#####################################################
\documentstyle[prd,tighten,preprint,aps]{revtex}

\newcommand{\square}{\kern1pt\vbox{\hrule height
1.2pt\hbox{\vrule width 1.2pt\hskip 3pt
   \vbox{\vskip 6pt}\hskip 3pt\vrule width 0.6pt}\hrule
height 0.6pt}\kern1pt}
\newcommand{\lsim}{\mbox{$\;\lower-.2ex
\hbox{$\textstyle<$}\;\!\!\!\!\!\!
\lower.7ex\hbox{$\textstyle \sim$}\;$}}
\newcommand{\gsim}{\mbox{$\;\lower-.2ex
\hbox{$\textstyle>$}\;\!\!\!\!\!\!
\lower.7ex\hbox{$\textstyle \sim$}\;$}}

\newenvironment{figcaption}[2]{
 \vspace{0.3cm}
 \refstepcounter{figure}
 \label{#1}
 \begin{center}
 \begin{minipage}{#2}
 \begingroup \small Fig. \thefigure: }{
 \endgroup
 \end{minipage}
 \end{center}}

\begin{document}

\begin{titlepage}
\baselineskip .15in
\begin{flushright}
WU-AP/64/96 \\
gr-qc/yymmddd \\
20th, December, 1996
\end{flushright}

\begin{center}
{\bf
{\large Gravitational Waves in Brans-Dicke Theory : \\
Analysis by Test Particles around a Kerr Black Hole}
}
\vskip 0.5cm
{\sc Motoyuki} SAIJO$^{\dag}$  ,
{\sc Hisa-aki} SHINKAI$^{\ddag}$  and  
{\sc  Kei-ichi} MAEDA$^{\S}$\\[1em]
{\em Department of Physics, Waseda University,
Shinjuku-ku, Tokyo 169, Japan~\dag,\S}\\
and\\
{\em Department of Phisics, Washington University,
St. Louis, MO63130-4899, USA~\ddag}
\end{center}
\begin{abstract}
Analyzing test particles falling into a Kerr black hole, we
study gravitational waves in Brans-Dicke theory of
gravity.
First we consider a test
particle  plunging with a constant azimuthal angle into a rotating
black hole and calculate  
the waveform and emitted energy  of both scalar and tensor
modes of  gravitational radiation. 
We find that the waveform as well as the energy of the scalar gravitational
waves weakly depends on the rotation parameter of black hole $a$ and 
on the azimuthal angle.  

Secondly, using a model of a non-spherical dust shell of test particles 
falling into a Kerr black hole,
we study when the scalar modes dominate.
When a black hole is rotating, the tensor modes 
do not vanish even for a ``spherically symmetric" shell, instead
a slightly oblate shell  minimizes their energy but with non-zero
finite value, which  depends on Kerr parameter $a$.
As a result, we find that the scalar modes dominate only 
for highly spherical collapse,
but they never exceed the tensor modes  unless
the Brans-Dicke parameter $\omega_{BD} \lsim 750 $ for 
$a/M=0.99$ or unless $\omega_{BD} \lsim 20,000 $ for 
$a/M=0.5$, where $M$ is mass of black hole.

We conclude that the scalar gravitational waves  with
$\omega_{BD} \lsim$ several thousands do not
dominate except for very limited situations (observation from the face-on
direction of a test particle falling into a Schwarzschild black hole or
highly spherical dust shell collapse into a Kerr black hole).  
 Therefore observation of polarization  is also required when we
determine the theory of gravity by the observation of gravitational
waves.

\end{abstract}

\begin{center}
December, 1996
\end{center}
\vfill
\dag~~Electronic mail : 696L0482@mn.waseda.ac.jp\\
\ddag~~Electronic mail :  shinkai@wurel.wustl.edu\\
\S~~Electronic mail :  maeda@mn.waseda.ac.jp\\
\end{titlepage}


\section{Introduction}

After the discovery of the binary pulsar PSR1916+13 by
Hulse and Taylor \cite{HT}, we believe in the existence of 
gravitational waves emitted from compact objects.  
Their continual observations over 20 years show that both 
the observed change rate of the orbital
period and that predicted by general relativity 
agree to within 0.3\% accuracy \cite{PSR1913}.

Currently, a number of worldwide projects for detecting
gravitational radiation directly using a km-scale  laser
interferometer such as LIGO, VIRGO, GEO600, and
TAMA are progressing.  In particular, the LIGO project \cite{LIGO}
will begin operating within a few years.
Among the many possible sources of gravitational waves, 
one expected plausible source  is 
a coalescing binary, composed of 
neutron star-neutron star (NS-NS), neutron star-black hole (NS-BH), 
or black hole-black hole (BH-BH).    
These direct detections  of gravitational waves will
enable us to see strong gravitational phenomena
such as coalescence of compact stars or black holes.  In order
to extract signals from noise,  we need to prepare in
advance a list of expected templates of 
gravitational waves, which depend
on many parameters of sources  (masses and spins of compact objects,
orbital angular momentum, eccentricity of their orbits, $\cdots$). 
Therefore one of the current most important subjects 
in general relativity is to
predict  a precise waveform of gravitational waves
for a set of given parameters. 

Furthermore, if we can directly observe gravitational waves, we
may also put some constraints on the theories of gravity, or
we might find some evidence for alternative theories of
gravity instead of general relativity.  
Although many alternatives have been proposed, most of them
have been rejected by experiments or observations except for a few
theories such as scalar tensor theories of gravity.  As the
simplest and proto-type among all scalar-tensor theories of
gravity, the Brans-Dicke (BD) theory
\cite{BD} of gravity is often considered as an alternative. 
We know the strictest bound for BD theory is
to the BD parameter $\omega_{BD} \gsim 500$ (the infinity limit
of $\omega_{BD}$ is just  general  relativity), which was made by
the observation of the Shapiro  time delay using Viking on Mars
\cite{VIKING}. The possibilities of testing gravity theory
using gravitational radiation as a tool
were also considered in the early days  \cite{wagonar70,eardley73},
but it was not effective in the weak gravitational field of the 
solar-system. 

Using the binary pulsar PSR1913+16, 
the consistency of general relativity was tested, 
as we mentioned above.
This binary  was also considered as a tool for checking other gravity
theories  \cite{eardley75,will77}.
Recent analysis by Will and Zaglauer \cite{WZ}, however,  concluded
that this source cannot be used  for testing BD theory
effectively.  This is because the two neutron stars
in the binary
are believed to have nearly equal masses, 
and the limit is also sensitive to the star models, 
hence the restriction on $\omega_{BD}$ was weak.
These  classical tests, however, are carried out
only in the weak gravitational field. 
On the contrary, observing gravitational waves directly from compact 
sources such as a BH-BH system will clarify some phenomena in strong
gravitational fields and may determine which theory of
gravity is correct. 
In order to analyze it from observational data, we also have to prepare
a list of templates of waveforms for other theories of gravity.  
In BD theory, gravitational waves have three modes, i.e.,  
a scalar mode, which we call a scalar gravitational wave (SGW), 
as well as two tensor modes  ($+$ and
$\times$ modes) of conventional gravitational waves, 
which we shall call tensor gravitational waves (TGWs) here.   
Will \cite{will94} studied a constrained bound of $\omega_{BD}$
from observing the inspiralling phase of compact binaries, and 
concluded that a bound strongly depends on mass difference between
components. 
To find the expected templates of SGWs, Shibata, Nakao and
Nakamura (SNN) \cite{SNN} calculated the waveform of  SGWs from
spherically symmetrical dust fluid collapse and concluded
that the advanced LIGO may detect SGWs even if the BD  coupling
constant $\omega_{BD}$ is $10^{4} \sim 10^{6}$.  

In this paper, we extend SNN's studies and figure out more
aspects of both SGWs and TGWs behavior in BD theory.  
Since a spherical symmetrical system is just an idealized model, 
if we consider a more realistic situation such as the case 
that spherically symmetry is no longer valid,
we may wonder whether SGWs can be really detected 
alongside TGW contributions.
If we can observe the signals with the full polarization information
of gravitational waves, then we can extract the scalar modes and do
not need to worry about whether the TGWs will dominate the SGWs.
However, if we have few observational bases, we may not be able
to separate the polarizations.  Hence it may be important to clarify
in which situation the SGWs can dominate.
Therefore, we  study 
gravitational waves in BD theory and compare
those scalar and tensor modes, analyzing   test particle motions
around a rotating black hole.
We analyze two cases: First, we calculate the
waveforms and emitted energy both of SGWs and of TGWs
 by a test particle plunging with a
constant azimuthal angle into a rotating black hole.    This 
linearized analysis using a test
particle  is known to have a reasonably good
agreement with fully relativistic numerical simulations in some 
cases such as a
head-on collision of two black holes \cite{APPSS}.  Our case, 
we believe, also
mimics a collision of two compact objects in BD theory.

 As for gravitational waves from gravitational collapse 
such as a supernova explosion followed by a formation of a neutron star, 
 we know
that no tensor modes are generated in a spherically symmetric case. 
Although the SGWs are still emitted in such a case and 
the theory of gravity might be determined by observing such SGWs, an
ansatz of spherical symmetry is  extremely idealized. 
Therefore, we next analyze a non-spherical dust shell collapsing into a
rotating black hole. Comparing the SGWs and TGWs,
we can see how much deviation from spherical symmetry  still gives
us the domination of the SGWs.  We then discuss the possibility of
identification of the SGWs without observing polarization.  

This paper is organized as follows. In \S2, 
we present our basic equations for calculating  gravitational radiation
in BD theory. 
We then analyze the waveforms and energy
both of the tensor and of the scalar modes 
 by a test particle plunging into
a rotating black hole in \S 3.   We also analyze  non-spherically
symmetric dust-shell collapse onto a Kerr black hole in \S 4. 
\S 5 is devoted to our conclusions and some remarks.  
Throughout this paper, we use the units of
$c=G_N=1$ \cite{footnote1} and the notations and definitions such as 
Christoffel symbols and curvature follow Misner-Thorne-Wheeler 
\cite{MTW}.

\section{Brans-Dicke Theory and Black Hole Perturbations}

   BD theory \cite{BD} is the simplest version of
scalar-tensor theories of gravity in which a scalar field
couples with metric tensor fields. The action $S$ of BD
theory is given by
\begin{equation}
S  =  \int d^{4}x \sqrt{-g} 
\left[ \frac{1}{16\pi} \left( \phi R -
\frac{\omega_{BD}}{\phi} g^{\mu \nu}
\nabla_\mu\phi\nabla_\nu\phi
\right)\right]  + S_{matter} ,
\label{eqn:action}
\end{equation}
where the coupling constant $\omega_{BD}$ is  called the BD
parameter. The theory is reduced to general relativity in the limit of
$\omega_{BD}
\rightarrow \infty$.  The conservative
observational bound on $\omega_{BD}$ is $\omega_{BD}
>500$, which is obtained from  the Shapiro time delay
experiment \cite{TEG}.

From the action (\ref{eqn:action}),
the field equations become
\begin{eqnarray}
G_{\mu \nu}   &=&  
 \frac{8\pi}{\phi} T_{\mu \nu} 
+ \frac{\omega_{BD}}{\phi^{2}} \left[(\nabla_{\mu} \phi) 
(\nabla_{\nu}\phi) -  \frac{1}{2}g_{\mu \nu} (\nabla\phi)^2
 \right] \nonumber \\
 && +  \frac{1}{\phi} ( \nabla_{\mu} \nabla_{\nu} \phi - 
g_{\mu \nu}
\square\phi ) ,
\label{eqn:BDgg} \\
\square \phi  & = &  {8\pi \over 3+2\omega_{BD}} T ,
\label{eqn:BD}
\end{eqnarray}
where $~G_{\mu \nu} = R_{\mu \nu} -
\frac{1}{2}g_{\mu \nu} R~$ is the Einstein tensor.  Before we
discuss perturbations of a black hole spacetime, it may be more 
convenient to introduce the Einstein frame, in which the metric and
scalar perturbations are decoupled. 
Using the conformal transformation from the physical
(Jordan-Brans-Dicke) frame to the Einstein frame as
\begin{equation}
g^{(E)}_{\mu \nu}=\left({\phi \over \phi_0} \right) ~ g_{\mu \nu},
 \label{conformal_transformation}
\end{equation}
the basic equations (\ref{eqn:BDgg}) and (\ref{eqn:BD}) are written as 
\begin{eqnarray}
G^{(E)}_{\mu \nu} & = &
8\pi  ( T^{(E)}_{\mu\nu} +  T^{(\Phi)}_{\mu\nu} )
  \label{eqn:Einstein} \\
\square^{(E)} \Phi & = & 2\left({\pi \over 
3+2\omega_{BD}}\right)^{1/2} T^{(E)}, 
\label{Eq_Phi}
\end{eqnarray}
where $\phi_0$ is a constant, which is fixed by our ansatz 
$G_N = 1$ \cite{footnote1}, and 
$\Phi$ is a  scalar field  in the Einstein frame defined by 
\begin{equation}
\Phi  = {1 \over 4} \left({3+2\omega_{BD} \over \pi}\right)^{1/2} \ln 
\left({\phi \over \phi_0}\right), \label{def:Phi}
\end{equation}
$G^{(E)}_{\mu \nu}$ is the Einstein tensor  defined by the metric
$g^{(E)}_{\mu \nu}$, and
\begin{eqnarray} 
T^{(E)}_{\mu\nu} & = & {1 \over \phi_0}
\exp\left[-4\left({\pi \over  3+2\omega_{BD}}\right)^{1/2}
\Phi \right] ~~ T_{\mu\nu} \\
T^{(E)} & = & {1 \over \phi_0}\exp\left[-8\left({\pi \over 
3+2\omega_{BD}}\right)^{1/2}
\Phi \right]~~  T \\
 T^{(\Phi)}_{\mu\nu} & = & \nabla_\mu \Phi\nabla_\nu \Phi -
{1\over 2} g^{(E)}_{\mu \nu}  (\nabla^{(E)}\Phi)^2 .
\end{eqnarray}     From these equations, 
it is easy to see that any vacuum solutions in
Einstein theory are equivalent with those in BD theory with
$\Phi$= constant.
In particular,  the Kerr solution with 
$\Phi=0$ is  a  solution  in BD theory as well.

We now discuss  
gravitational waves emitted by  a  test particle plunging into a Kerr
black hole.  In order to calculate the waveform of emitted gravitational radiation and its energy,
 we perturb the Kerr background spacetime.  Since we are
interested in a test particle motion, we adopt  the matter Lagrangian
as  
\begin{equation}
S_{matter}=\int d\tau  \mu
(\phi) ,
\end{equation}
where $\tau$ and $\mu$ are   proper time and mass of a 
test particle, respectively. 
In the case that the 
self-gravity of the test particle cannot be ignored,
for example an extrapolation of a test particle into 
a neutron star or a black hole,
$\mu$ should depend on the BD scalar $\phi$, which
does not guarantee that the test particle moves along the
geodesic and the gravitational weak equivalence principle
is no longer valid (Nordtvert effect).  In this paper, however,
we consider only the case of $\mu$=constant just for
simplicity.  
When we extrapolate our results into the case of the direct
collision of two compact objects, we should  take into
account the $\phi$-dependence in $\mu$ \cite{eardley75},
but the difference may not be so
large unless the test particle is a black hole, in which case
scalar gravitational waves will not be emitted, at least within
the linear perturbation approximation \cite{BH-s}.
The energy-momentum
tensor  of the test particle  is now
\begin{equation}
T^{\mu \nu}  =  \frac{\mu}{\sqrt{-g}} \int  d\tau
\frac{dz^{\mu}}{d\tau} 
\frac{dz^{\nu}}{d\tau} \delta^{(4)}(x-z(\tau)) , 
\end{equation}
where $z^\mu (\tau)$ is the geodesic of a test particle.

Now we linearize our basic equations in the Einstein frame around a
background vacuum spacetime :
\begin{equation}
g^{(E)}_{\mu \nu} = g^{(0)}_{\mu \nu} + h^{(E)}_{\mu \nu}  ~~~~{\rm
and}~~~~~
\Phi = \Phi^{(0)} + \Phi^{(1)},
\end{equation}
where $\Phi^{(0)} = 0$.
Then the first order perturbation equations are  
\begin{eqnarray}
G^{(1)}_{\mu\nu}[h^{(E)}_{\alpha \beta}] 
& = &
8\pi T_{\mu\nu}^{(E)}[\mu, g^{(0)}_{\mu
\nu}, \Phi^{(0)}=0] 
\label{eqn:perturbation_Einstein}\\
\square^{(0)} \Phi^{(1)} & = &  2\left({\pi \over 
3+2\omega_{BD}}\right)^{1/2}
 T^{(E)}[\mu, g^{(0)}_{\mu
\nu}, \Phi^{(0)}=0]
  \label{eqn:perturbation_Phi},
\end{eqnarray}
where 
\begin{eqnarray}
G^{(1)}_{\mu\nu}[h^{(E)}_{\alpha \beta}]  & = &
{1 \over 2}\left[-\nabla^{(0)}_\mu \nabla^{(0)}_\nu h^{(E)}
-\nabla^{(0)}_\alpha \nabla^{(0) \alpha} h^{(E)}_{\mu\nu}
+\nabla^{(0)}_\nu \nabla^{(0) \alpha} h^{(E)}_{\alpha\mu}
+\nabla^{(0)}_\mu \nabla^{(0) \alpha} h^{(E)}_{\alpha\nu}
\right. \nonumber \\
&& ~~~\left. -g^{(0)}_{\mu
\nu} \left(-\nabla^{(0)}_\alpha \nabla^{(0) \alpha}h^{(E)}
+\nabla^{(0) \alpha} \nabla^{(0) \beta}h^{(E)}_{\alpha\beta}\right)
\right]
\end{eqnarray}
 is the first order  perturbation of the Einstein tensor
$G_{\mu\nu}$ and
$\square^{(0)} $ denotes the D'Alembertian of the background
spacetime.  Since mass of the test particle $\mu$ is a first order
quantity, both energy-momentum tensors in Einstein frame and in
Jordan-Brans-Dicke frame differ only by a constant factor $1/\phi_0$ in the linear
perturbation equations (\ref{eqn:perturbation_Einstein}),
(\ref{eqn:perturbation_Phi}).  We also find
that the metric perturbation equation
(\ref{eqn:perturbation_Einstein})  for
$h^{(E)}_{\mu \nu}$  is exactly the
same as those in Einstein theory \cite{footnote2}. 

In the 
Jordan-Brans-Dicke frame, the perturbation variables,  which are expanded as
\begin{equation}
g_{\mu \nu} = 
g^{(0)}_{\mu \nu} + h_{\mu \nu} ~~~~{\rm and}~~~~~
\phi = \phi_0 + \phi^{(1)}, 
\end{equation}
are given by the variables in the Einstein frame as 
\begin{equation}
h_{\mu \nu} = h^{(E)}_{\mu \nu} -  g^{(0)}_{\mu
\nu}{\phi^{(1)} \over 
\phi_0}  ~~~~{\rm and}~~~~~
{\phi^{(1)} \over 
\phi_0} = 4\left({\pi \over  3+2\omega_{BD}}\right)^{1/2}
\Phi^{(1)} .
\end{equation}
 For the equation for 
$\phi^{(1)}$,  from Eq. (\ref{eqn:perturbation_Phi}) or (\ref{eqn:BD}), we
obtain
\begin{equation}
\square^{(0)} \phi^{(1)} ={8\pi \over  3+2\omega_{BD}} T
\label{Eq_phi}.
\end{equation}
As for the metric perturbations in the Jordan-Brans-Dicke frame,
we just have to study the perturbations and
gravitational waves in Einstein theory,
since the basic
equations in the Einstein frame are exactly the same as those in the
Einstein theory,  which we have better knowledge in treating in
ordinary procedures.

\section{Gravitational Waves from Test Particles Falling  
into Kerr Black Hole}
\subsection{Metric Perturbations}

We start considering a test particle motion 
 in the  Kerr background spacetime and analyze first 
the tensor modes of its gravitational
waves (TGWs) emitted by a plunging test particle.  
In order to use these results in the analysis in \S 4, 
we take into account a test particle not only moving in the equatorial plane
but also moving off the equatorial plane,  which results have not
 been obtained so far.
This means that  we
will also present new results in Einstein theory, i.e., the 
waveforms and energy flux in the case of test particles falling
with a constant  azimuthal angle into a Kerr black hole.

Let us begin from a brief review
of a black hole perturbation analysis  in Einstein theory,
which gives the same results for tensor modes in BD theory up to a factor $1/\phi_0$.  
In order to calculate metric perturbations of a Kerr spacetime in the
Einstein theory,  we adopt the Sasaki-Nakamura formalism \cite{SN}, in 
which 
the Regge-Wheeler equation is generalized by a transformation from the 
Teukolsky equation \cite{BPT}.  
The radial wave equation becomes
\begin{eqnarray}
\left[ \frac{d^{2}}{dr^{*2}} - F(r)\frac{d}{dr^*} - U(r)  \right]
X_{lm\omega}(r) & = & S_{lm\omega}(r),
\label{eqn:GeneRW}
\end{eqnarray}
where  the tortoise coordinate 
$r^{*}$ is
defined as
\begin{equation}
dr^* = {r^2+a^2 \over \Delta} dr.
\end{equation}
with $\Delta  =  r^{2} - 2 M r +a^{2}$.
The potential functions $F(r)$ and $U(r)$ are given as
\begin{eqnarray*}
F(r) & = & \frac{\Delta}{r^{2}+a^{2}}\frac{1}{\gamma}
\frac{d\gamma}{dr}, 
\\
U(r) & = & G^{2} - FG + 
\frac{\Delta} {(r^{2}+a^{2})} \frac{dG}{dr}
+ \frac{\Delta }{(r^{2}+a^{2})^{2}} U_{1} ,
\end{eqnarray*}
with
\begin{eqnarray*}
\gamma (r) & = & c_{0}+\frac{c_{1}}{r}+\frac{c_{2}}{r^{2}}
+\frac{c_{3}}{r^{3}}+
\frac{c_{4}}{r^{4}},\\
 & & c_{0} = -12iM\omega + \lambda(\lambda+2)
+12a\omega(m-a\omega),\\
 &  &  c_{1} = 8ia[3a\omega -
\lambda(m-a\omega)],\\ 
 & & c_{2}= 24iaM(m-a\omega) +
12a^{2}[1-2(m-a\omega)^{2}],\\ 
&  & c_{3} =
24ia^{3}(m- a\omega)-24Ma^{2},\\
& &  c_{4}  = 12a^{4},\\
G(r) & = & -\frac{1}{r^{2}+a^{2}}{d\Delta \over dr} + \frac{r\Delta}{(r^{2}
+a^{2})^{2}},\\ 
U_{1} (r) & = & V_{T} +
\frac{\Delta^{2}}{\beta}\left[ 
\frac{d}{dr}\left(2\alpha+\frac{d\beta /dr}
{\Delta}\right)-\frac{d\gamma /dr}{\gamma}\left(\alpha + 
\frac{d\beta /dr}{\Delta}\right)\right],\\
& & \alpha  =-i\frac{K\beta}{\Delta^{2}}+3i\frac{dK}{dr}+\lambda
+\frac{6\Delta}{r^{2}},\\
&  & \beta  = 2\Delta\left( -iK+r-M-\frac{2\Delta}{r} \right).
\end{eqnarray*}
$V_{T}$ is the potential in the Teukolsky equation, which is given as 
\begin{equation}V_{T}(r)  =  - \frac{K^{2} +4 i (r-M) K}{\Delta} 
+ 8 i\omega r +
\lambda ,
\end{equation}
 where $K=(r^2+a^2)\omega -ma$.
In Appendix A, we present the explicit form of the source term 
$S_{lm\omega}(r)$
for the general motion of a test particle.

 To describe the wave
function $X_{lm\omega}(r)$ of  Eq. (\ref{eqn:GeneRW}) using the
Green function method,  we need two independent homogeneous
solutions of the Sasaki-Nakamura equation, i.e., 
\begin{eqnarray*}
X_{lm\omega}^{in(0)}(r) & = & \left\{
\begin{array}{ll}
e^{-i k r^{*}} & r^{*}\rightarrow -\infty \\
A_{lm\omega}^{in}e^{-i \omega r^{*}} + A_{lm\omega}^{out}e^{i 
\omega r^{*}} & r^{*}\rightarrow
\infty 
\end{array}
\right. ,\\
X_{lm\omega}^{out(0)}(r) & = & \left\{
\begin{array}{ll}
B_{lm\omega}^{in}e^{-i k r^{*}} + B_{lm\omega}^{out}e^{i k r^{*}}
& r^{*}\rightarrow - \infty \\
e^{i \omega r^{*}} & r^{*}\rightarrow \infty 
\end{array}
\right. ,
\end{eqnarray*}
where $k=\omega-ma/2r_{+}$ with $r_{+} = M+\sqrt{M^2-a^2}$ being the 
radius of
the  event horizon,  and the Wronskian $W^{\rm (TGW)}$
\begin{eqnarray*}
W^{\rm (TGW)}  \equiv X_{lm\omega}^{in(0)}  \frac{d
X_{lm\omega}^{out(0)}}{dr^{*}} - X_{lm\omega}^{out(0)}
\frac{dX_{lm\omega}^{in(0)}}{dr^{*}} = 2i\omega
A_{lm\omega}^{in}\frac{\gamma}{c_{0}}.
\end{eqnarray*}
Then the inhomogeneous solution of Eq. (\ref{eqn:GeneRW}) becomes
\begin{eqnarray*}
X_{lm\omega}(r) & = & 
X_{lm\omega}^{in(0)} 
\int_{r^{*}}^{\infty}  {S_{lm\omega}(r)\over {W^{\rm (TGW)}}}
X_{lm\omega}^{out(0)}~dr^{*} + X_{lm\omega}^{out(0)}
\int_{-\infty}^{r^{*}} {S_{lm\omega}(r)\over {W^{\rm (TGW)}}}
X_{lm\omega}^{in(0)}~dr^{*}.
\end{eqnarray*}
 For observation, we only have to know the asymptotic behavior  of the 
wave function 
$X_{lm\omega}(r)$ at infinity,
which is 
\begin{eqnarray}
X_{lm\omega}^{(\infty)}(r) & = & 
A_{lm\omega}^{(\infty)} e^{i\omega r^{*}}, 
\label{eqn:RWsol}
\end{eqnarray}
where 
\begin{equation}
A_{lm\omega}^{(\infty)} = \frac{c_0}{2i\omega 
A_{lm\omega}^{in}} 
\int_{-\infty}^{\infty}  {S_{lm\omega}(r)\over \gamma}
X_{lm\omega}^{in(0)}(r)~dr^{*} .
\end{equation}
Then the waveform  at infinity is 
\begin{eqnarray}
h_{+}-ih_{\times} & = & \frac{8}{r} \int_{-\infty}^{\infty} 
d\omega~e^{i\omega (r^{*}-t)} \sum_{l,m}  {A_{lm\omega}^{(\infty)}
\over c_0}
~\mbox{}_{-2}S_{lm}^{a\omega}(\theta) ~{e^{im\varphi} \over
\sqrt{2\pi}},
\label{eqn:TGW}
\end{eqnarray}
where $\mbox{}_{s}S_{lm}^{a\omega} (\theta)$ is a
spin-weighted  spheroidal function, which obeys
\begin{eqnarray}
\lefteqn{\left[ \frac{1}{\sin\theta} \frac{d}{d\theta} 
\left( \sin\theta \frac{d}{d\theta} \right) - \Biggl( 
a^{2}\omega^{2}\sin^{2}\theta \right.} \nonumber \\
&& \left.
+ \frac{(m+s\cos\theta)^{2}}{\sin^{2}\theta}
+ 2a\omega s\cos\theta-s ~-2am\omega-\lambda  \Biggr) \right] ~
\mbox{}_{s}S_{lm}^{a\omega}(\theta) 
=  0,
\end{eqnarray}
with $s=-2,-1,0,1$ or $2$ and $\lambda$ being the eigenvalue.
$\mbox{}_{s}S_{lm}^{a\omega} (\theta)$  is normalized as
\begin{eqnarray*}
\int_{0}^{\pi} d\theta \sin\theta 
|\mbox{}_{s}S_{lm}^{a\omega}(\theta)|^{2} 
& = & 1.
\end{eqnarray*}  From Eq. (\ref{eqn:TGW}), the energy flux of 
gravitational waves at infinity becomes
\begin{eqnarray}
\frac{dE^{\rm (TGW)}}{dt} & = & \frac{r^2}{32\pi} \int {\partial h^{ij}_{TT}
\over \partial t}
{\partial h^{ij}_{TT}
\over \partial t}  d\Omega  \nonumber \\
& = & \frac{4}{\pi} \sum_{l,m} \left| \int_{-\infty}^{\infty} d\omega
~\omega  \frac{A_{lm\omega}^{(\infty)}}{c_{0}} e^{i\omega(r^{*}-t)} \right|^{2}.
\end{eqnarray}
Then, the total energy and the energy spectrum of the TGW are
given as 
\begin{eqnarray}
E^{\rm (TGW)} & = & \int^{\infty}_{-\infty} d\omega~\sum_{l,m}  \left(
\frac{dE^{\rm (TGW)}}{d\omega}
\right)_{lm\omega} , 
\end{eqnarray}
and 
\begin{equation}
\left( \frac{dE^{\rm (TGW)}}{d\omega} \right)_{lm\omega} = 8
\omega^{2}\left| \frac{A_{lm\omega}^{(\infty)}}{c_{0}}\right|^{2} ,
\label{eqn:GravEnergy}
\end{equation}
respectively. 

Solving this equation, Kojima and Nakamura \cite{KN}
analyzed the gravitational radiation induced by 
a test particle with mass $\mu$ and
angular momentum $L_{z}$  plunging on the equatorial plane or along  the
rotation axis  into a Kerr black hole with mass
$M$ and angular momentum
$J=Ma$.  They calculated the waveform (only
$h_{+}$ mode) and the emitted energy. 
They  pointed out that  the behaviors are  similar to those of
a Schwarzschild black hole but the amplitude  becomes larger as the
rotation parameter $a$ gets larger. They showed that the emitted energy from 
the particle moving along the axis is smaller than that from the particle
moving on the equatorial plane. They also calculated the total energy,  linear
momentum and  angular  momentum extracted from the system by gravitational
radiation and showed that all of them become  larger as the
rotation of black hole gets faster.  

For TGWs in BD theory, we have to multiply $1/\phi_0$ by the energy
momentum  tensor, and then the energy flux.  But since $1 \leq \phi_0
< 1.001$  for $\omega_{BD}>500$, we may ignore such a factor here.
In order to compare the tensor modes with the scalar mode, we 
calculate the  waveform and emitted energy of gravitational
radiation induced by a test particle plunging with a constant
azimuthal angle into a  Kerr black hole.  The particle motion with a
constant azimuthal angle is a geodesic in Kerr spacetime, if
$E=\mu$ and $L_z=0$, which we assume here (see Appendix).
  We have analyzed such a
special  geodesic because in the next section we will use it 
in the analysis of
non-spherical dust shell collapse . We wish to emphasize
that the results with the azimuthal angle
$\theta \neq 0$ or $\neq
\pi/2$  are new even in Einstein theory \cite{footnote3}.  We have 
checked
our numerical codes with  the latest results, Mino, Shibata and
Tanaka \cite{MST}.  

 We show the emitted waveforms and emitted energy for several values of the
Kerr parameter $a$ ($a=0,~0.5,~0.9,~0.99$) and of the azimuthal angle $\theta$
($\theta = \pi/6, \pi/3, \pi/2$ (the equatorial plane))
 in  Figs. 1 and 2.  
We assume that the observer is in the position of 
$\theta=\pi/2,~\phi=0$.  
One of the important results is that 
only in the case of an  observer perpendicular to the direction of a
direct collision,  and  for very small rotation of compact objects, 
do we find little gravitational radiation (Fig. 1 $a=0$ case).  This  is consistent with a
full numerical analysis of two black hole collision 
\cite{APPSS}.

We also present the total energy of TGWs in terms of the azimuthal
angle in Fig. 2.
The emitted energy changes smoothly in terms of the azimuthal angle from the
largest one for
$\theta=\pi/2$ to the smallest one for $\theta=0$, each values  are
found in 
\cite{KN} and \cite{NS}.

\subsection{Scalar Perturbations}

Next we consider the scalar mode $\phi^{(1)}$ of the perturbation. 
We  expand $\phi^{(1)}$ using spin-weighted spheroidal function as 
\begin{eqnarray}
\phi^{(1)} & = & \sum_{l,m} \int d\omega~e^{-i\omega t}
\frac{Z_{lm\omega}(r)} {\sqrt{r^{2}+a^{2}}}
~\mbox{}_{0}S_{lm}^{a\omega}(\theta) ~{e^{im\varphi} \over
\sqrt{2\pi}},
\label{eqn:phi} 
\end{eqnarray} 
then, from the field equation (\ref{Eq_phi}), the radial wave function 
$Z_{lm\omega}(r)$ obeys the  differential equation
\begin{eqnarray}
\left[ \frac{d^{2}}{dr^{*2}} +  V(r) \right]
Z_{lm\omega}(r) & = & \frac{8
\pi}{2\omega_{BD}+3}
\frac{\Delta}{(r^{2}+a^{2})^{3/2}}
T_{lm\omega}(r), 
\label{eqn:ScalarBasic}
\end{eqnarray}
where the potential $V(r)$ is given by
\begin{equation}
 V(r) = \left(
\omega-\frac{am}{r^{2}+a^{2}}\right)^{2} -
\frac{\Delta[\lambda(r^{2}+a^{2})^{2}+(2Mr^{3}
+a^{2}r^{2}-4Ma^{2}r+a^{4})]} {(r^{2}+a^{2})^{4}},
\end{equation}
and  the trace of energy momentum tensor  is expanded as 
\begin{eqnarray}
\Sigma T & = & \sum_{l,m} \int d\omega~e^{-i\omega t} 
T_{lm\omega}(r)
~\mbox{}_{0}S_{lm}^{a\omega}(\theta) ~{e^{im\varphi} \over
\sqrt{2\pi}},
\end{eqnarray}
with $\Sigma = r^2 +a^2 \cos^2 \theta$.  We find that
\begin{eqnarray}
T_{lm\omega}(r) & = & 
-{1 \over (2\pi)^{3/2}}\left|{d\tau(r) \over dr}\right|
 e^{i(\omega t(r)-m \varphi (r))} 
~\mbox{}_{0}S_{lm}^{a\omega}(\theta (r)),
\end{eqnarray}
where we have to insert a test particle orbit 
described in terms of $r$, i.e., 
$z^\mu =(t(r), r, \theta (r), \varphi (r))$.
We again use the Green function method to derive the wave function
$Z_{lm\omega}(r)$. Defining two independent homogeneous solutions as
\begin{eqnarray*}
Z_{lm\omega}^{in(0)}(r) & = & \left\{
\begin{array}{ll}
e^{-i k r^{*}} & r^{*}\rightarrow -\infty \\
C_{lm\omega}^{in}e^{-i \omega r^{*}} + C_{lm\omega}^{out}e^{i 
\omega r^{*}} & r^{*}\rightarrow
\infty 
\end{array}
\right. ,\\
Z_{lm\omega}^{out(0)}(r) & = & \left\{
\begin{array}{ll}
D_{lm\omega}^{in}e^{-i k r^{*}} + D_{lm\omega}^{out}e^{i k r^{*}}
& r^{*}\rightarrow - \infty \\
e^{i \omega r^{*}} & r^{*}\rightarrow \infty 
\end{array}
\right. ,
\end{eqnarray*}
 and the Wronskian $W^{\rm (SGW)}$
\begin{eqnarray*}
W^{\rm (SGW)}  \equiv   Z_{lm\omega}^{in(0)}  \frac{d
Z_{lm\omega}^{out(0)}}{dr^{*}} - Z_{lm\omega}^{out(0)}
\frac{dZ_{lm\omega}^{in(0)}}{dr^{*}} = 2i\omega
C_{lm\omega}^{in} , 
\end{eqnarray*}
 the inhomogeneous solution of Eq. (\ref{eqn:ScalarBasic}) is given as 
\begin{eqnarray*}
Z_{lm\omega}(r) & = & \frac{8\pi}{2\omega_{BD}+3} \frac{1}{W^{\rm
(SGW)}} 
\left( Z_{lm\omega}^{in(0)} 
\int_{r^{*}}^{\infty}  \frac{\Delta}{(r^{2}+a^{2})^{3/2}}
T_{lm\omega}(r) Z_{lm\omega}^{out(0)}~dr^{*} \right. \nonumber\\
&&\left. + Z_{lm\omega}^{out(0)}
\int_{-\infty}^{r^{*}} \frac{\Delta}{(r^{2}+a^{2})^{3/2}}
T_{lm\omega}(r) Z_{lm\omega}^{in(0)}~dr^{*} \right).
\end{eqnarray*}
We only need the asymptotic behavior  of the 
wave function 
$Z_{lm\omega}(r)$ at infinity for observation,
which is 
\begin{eqnarray}
Z_{lm\omega}^{(\infty)}(r) & = & 
C_{lm\omega}^{(\infty)} e^{i\omega r^{*}}, 
\label{eqn:?}
\end{eqnarray}
where 
\begin{equation}
C_{lm\omega}^{(\infty)} = \frac{1}{2i\omega 
C_{lm\omega}^{in}} 
\int_{-\infty}^{\infty} \frac{\Delta}{(r^{2}+a^{2})^{3/2}}
T_{lm\omega}(r) Z_{lm\omega}^{in(0)}(r)~dr^{*} .
\end{equation}
Then the waveform  at infinity is derived as 
\begin{eqnarray}
\phi^{(1)} & = & \frac{1}{\sqrt{r^{2}+a^{2}}} \int_{-\infty}^{\infty} 
d\omega~e^{i\omega (r^{*}-t)} \sum_{l,m}  C_{lm\omega}^{(\infty)}
~ \mbox{}_{0}S_{lm}^{a\omega}(\theta) ~{e^{im\varphi} \over
\sqrt{2\pi}}.
\label{eqn:SGW}
\end{eqnarray}

The energy flux of the SGW observed at infinity is defined as \cite{TEG}
\begin{equation}
\frac{dE^{\rm (SGW)}}{dt} = \frac{(2\omega_{BD}+3)^{2}}{32\pi(\omega_{BD}+2)} \int 
 \left({\partial\phi\over \partial t}\right)^2 dS
\label{eqn:ScalarEnergy_def}.
\end{equation}
Then the total energy $E^{\rm (SGW)}$ is now 
\begin{equation}
E^{\rm (SGW)}
= \sum_{l,m}  \int_{-\infty}^{\infty} d\omega
\left( \frac{dE^{\rm (SGW)}}{d\omega} \right)_{lm\omega} 
\label{eqn:ScalarEnergy},
\end{equation}
with
\begin{eqnarray}
\left( \frac{dE^{\rm (SGW)}}{d\omega} \right)_{lm\omega} 
& = & \frac{(2\omega_{BD}+3)^{2}}{16(\omega_{BD}+2)}
\omega^{2} |C_{lm\omega}^{(\infty)}|^{2}.
\label{eqn:ScalarEnergy_spectrum}
\end{eqnarray}

In order to check our numerical code, for the case of a test particle
falling into Schwarzschild background, we have compared our 
results with those by Shibata, Nakao and Nakamura \cite{SNN}, finding a
good agreement.  
We show our results in Figs. 3 to 6 for 
$a$=0, 0.5, 0.9, and 0.99 and
for $\theta = \pi/6, \pi/3$ and $\pi/2$.  Here we assume
$\omega_{BD}=500$.  Since $\omega_{BD} \gg 1$, the amplitude and
energy of  the
SGWs are both approximately proportional to 
$1/\omega_{BD}$ (see Eqs. (\ref{eqn:ScalarBasic}) and 
(\ref{eqn:ScalarEnergy_spectrum})) and then we 
can rescale our
results for any arbitrary value of  $\omega_{BD}$.

 From the waveforms (Fig. 3), energy spectrum (Fig.4) and
the total energy  (Fig. 5), we find that the results 
depend only slightly on the values of 
$a$ and $\theta$.  This $\theta$ independence may be understood from 
the fact that the $s$ wave ($l=0$) is dominant in the SGWs.  
As for the Kerr parameter, although the phase difference depends on $a$, we find that the waveforms themselves little depends on $a$.
The energy spectrum has three main peaks (see Fig. 4) and the
frequency of the largest peak is almost the same as that of the quasi-normal
mode  (QNM).  These QNMs of scalar
 modes ($l=0$) are obtained by Simone-Will \cite{simone-will} and Andersson
\cite{niles92} for the Schwarzschild black hole case, using WKB
approximation and  phase-integral method, respectively.  Here,
using Leaver's  technique \cite{Leaver} of continued fractions, 
we derive  the
$l=0$ modes of QNMs for a Kerr black hole and try to fit the tail part of  the
obtained waveform by them. 
In Fig. 6, we show both the calculated waveform (bold line)
and the curve fitted by QNM (dotted line).
The results show that the ringing tail of the SGW can also fit by the
QNM.   To draw a fitting curve, we need only the first QNM, i.e.,
\begin{eqnarray}
\phi^{(1)} & = & {A\mu \over r} e^{{\rm Im}[\omega_{1}]
(t-r^{*}+b)} 
\cos({\rm Re}[\omega_{1}] (t-r^{*}+b)).
\label{eqn:BDcurve}
\end{eqnarray}
with $A=-0.0012,~b=-33,~M\omega_1=0.11045 -  0.10490 i$ for
$a=0$, 
$A=0.0013, b=-39, M\omega_1=0.11251 - 0.10000 i$ for $a=0.5M$,
$A=-0.0014,~b=-72,~ 
M\omega_1=0.11380 -  0.09160 i$ for $a=0.9M$.
However,the QNM fit to the ringing tail of the SGW is not as good as
the fit to the TGWs.
The situation does not change even if we
include the higher QNMs.   This may be because 
we cannot separate the burst part and the ringing tail part of SGWs,
we cannot fit the tail part of SGW only by the QNMs.

Because of the ratio of the maximum amplitude of GW  to
that of SGW is 200:1, a detection
of the SGW itself is usually very difficult, unless we observe the
polarizations and divide the tensor and the scalar modes. 
Only one exceptional case is that the observer's line
of sight is in the direction of a falling test particle into a 
Schwarzschild
black hole.   In this case,  
the SGW always dominates the tensor modes.

\section{Gravitational Waves from Non-spherical Dust Shell Collapse}

As we found in the previous section, 
when a test particle is falling into a Kerr black hole, 
the ratio of maximum amplitude of SGW to TGW is 1:200. 
This suggest that direct detection of SGWs from a collision of compact bodies 
is a hard task. 
However, the fact that SGWs are generated even in a spherically
symmetric system, while TGWs require a deviation from spherical 
symmetry,  gives us an expectation that
the SGW may become dominant in some situations such as an almost spherical
 supernova explosion.  
Since a completely spherically symmetric system is
unrealistic, 
we may ask how much deviation is allowed for the SGW to dominate TGW. 
In this section, we analyze the gravitational radiation from a 
collapsing  non-spherical dust-shell, which consists of test particles  (so its self
gravity is  ignored),  into a
Kerr black hole. In early studies by Nakamura and Sasaki \cite{NS} and Shapiro and Wasserman \cite{shapiro-shell-infall}, the TGWs for a deformed
shell falling into a Schwarzschild black hole were investigated. 
Our study is an extension of their researches to a rotating black
hole case and/or to BD theory.

The gravitational radiation from a dust shell is derived from a linear 
combination of test particles falling into a black hole.  
If the dust shell is  spherically
symmetric, a phase cancellation mechanism reduces the energy of 
gravitational waves to zero in the Schwarzschild black hole case.

Let us begin by defining the shape of a dust shell by a function
$r = r_{i}(\theta)~(i=1, 2)$;
\begin{eqnarray}
r_{1}(\theta) & = & r_{0} (1+\delta\cos^{2}\theta) \hspace{5mm} 
(\delta\geq 0)\hspace{5mm} ({\rm prolate}), \\
r_{2}(\theta) & = & r_{0} (1+\delta\sin^{2}\theta) \hspace{5mm} 
(\delta\geq 0)\hspace{5mm} ({\rm oblate}), 
\end{eqnarray}
where $\delta$ is a parameter which denotes a deviation from a 
spherically
symmetric dust shell and $r_{0}$ is a typical radius of the shell. 

In the case of a Schwarzschild black hole, test particle motion with 
zero angular momentum does 
not depend on the azimuthal angle $\theta$, and the dynamics of particles in 
 the shell can be replaced by that on
   the equatorial plane.
This fact makes our calculation of energy of gravitational waves easy as follows.
  Using 
the shape function $r_i(\theta)$, we introduce a correction function 
$f_{lm\omega}$ \cite{NS,HSW} as 
\begin{eqnarray}
f_{lm\omega}  & = & \frac{1}{2} \int_0^\pi d\theta
\sin\theta ~e^{i\omega [t(r_{0})-t(r_{i}(\theta)) ]}  P_{l}(\theta)
\delta_{m0},
\label{eqn:fS} 
\end{eqnarray}
where $P_{l}$ is the  Legendre  function and $t(r)$ is the coordinate
time 
 when the particle comes to a position  $r$.  
Then we define the source  term
$S_{lm\omega}$ and  wave function $X_{lm\omega}$ for non-spherical
dust  collapse from those of test particles   as
\begin{eqnarray}
S^{\rm [shell]}_{lm\omega} & = & f_{lm\omega} S_{lm\omega}^{\rm 
[particle]},\\ 
X_{lm\omega}^{\rm [shell]} & = & f_{lm\omega}
X_{lm\omega}^{\rm [particle]}.
\end{eqnarray}
We then obtain  a total energy of gravitational
radiation  from   non-spherical dust shell collapse as
\begin{eqnarray}
E^{\rm (TGW)} & = & \sum_{l,m} \int_{-\infty}^{\infty} d\omega \left(
\frac{dE^{\rm (TGW)}}{d\omega} 
\right) _{lm\omega} 
\label{eqn:TotalE}
\end{eqnarray}
where 
\begin{eqnarray}
\left( \frac{dE^{\rm (TGW)}}{d\omega} 
\right) _{lm\omega} &=& |f_{lm\omega}|^{2}
\left( \frac{dE^{\rm (TGW)}}{d\omega} \right)_{lm\omega}^{\rm
[paticle]} 
 \nonumber \\
& = & 
8 \omega^{2} |f_{lm\omega}|^{2}
|\frac{X_{lm\omega}^{(\infty) [particle]} }{c_{0}}|^{2} .
\label{eqn:zureE}
\end{eqnarray}
As for the SGW, we have 
\begin{eqnarray}
E^{\rm (SGW)} & = & \sum_{l,m} \int_{-\infty}^{\infty} d\omega \left(
\frac{dE^{\rm (SGW)}}{d\omega} 
\right) _{lm\omega} 
\label{eqn:TotalE_SGW}
\end{eqnarray}
where 
\begin{eqnarray}
\left( \frac{dE^{\rm (SGW)}}{d\omega} 
\right) _{lm\omega} &=&
\frac{(2\omega_{BD}+3)^{2}}{16(\omega_{BD}+2)} \omega^{2} |f_{lm\omega}|^{2}
|Z_{lm\omega}^{(\infty) [particle]} |^{2} .
 \label{eqn:zureE_SGW}
\end{eqnarray}

In a Kerr spacetime, however, we cannot adopt this simple formula
because   the radial wave function depends on the azimuthal angle
$\theta$, which means that we cannot factorize a correction function
as Eq. (\ref{eqn:zureE}).  In this case, we have to
calculate the  total energy by  summing up test particles. 
Therefore,  as for test particles of the dust shell, 
we choose  $E=\mu$ and $L_z=0$, for which  a trajectory
of $\theta$=constant becomes a geodesic.
First, using Eq. (\ref{eqn:GravEnergy}),
we prepare the wave function from a  test  particle 
with constant $\theta$, which we denote  $X_{lm\omega}(r;\theta)$.  
Then we calculate the energy spectrum of the TGW as 
\begin{eqnarray}
\left( \frac{dE^{\rm (TGW)}}{d\omega} \right)
_{lm\omega} & = & 2\omega^2 \left|  \int_{0}^{\pi} d\theta\sin\theta  
\frac{X_{lm\omega}^{(\infty)}(r;\theta) }{c_{0}}~e^{i\omega
[t(r_{0})-t(r_{i}(\theta))]}
 \right|^{2} \delta_{m0},
\label{eqn:zuredEK}
\end{eqnarray}
where $X_{lm\omega}^{(\infty)}(r;\theta)$ is the asymptotic form of
$X_{lm\omega}(r;\theta)$ at infinity.
To evaluate the
integration  (\ref{eqn:zuredEK}), we prepare data at  60 points for $\theta$ in
$[0,\pi]$. The total energy is given by Eq. (\ref{eqn:TotalE}).

As for the SGW, we also find
\begin{eqnarray}
\left( \frac{dE^{\rm (SGW)}}{d\omega} 
\right) _{lm\omega} &=& \frac{(2\omega_{BD}+3)^{2}}{16(\omega_{BD}+2)} 
\nonumber \\
&~& \times
\omega^2 \left|  \int_{0}^{\pi} d\theta\sin\theta  
Z_{lm\omega}^{(\infty)}(r;\theta) ~e^{i\omega [t(r_{0})
-t(r_{i}(\theta))]}
 \right|^{2} \delta_{m0}.
 \label{eqn:zureS}
\end{eqnarray}

We show our results in Figs. 7 and 8. 
For a
Schwarzschild background,
 Nakamura and Sasaki \cite{NS}  calculated the total 
energy of TGWs from a dust shell and pointed out that the energy will
be  generated maximally for a  deviation of $\delta_{cr}^{\rm (TGW)} \sim 0.5$
from spherical 
symmetry.  The reason is that if the dust shell is close to a spherical
symmetry, the TGW will not be generated so much while if the deviation is 
larger
than $\delta_{cr}^{\rm (TGW)}$, a phase cancellation effect will also reduce the emitted
energy \cite{NS} .

For the case of  SGW, we find that the energy does not depend on the 
deformation parameter $\delta$ if  $\delta$ is smaller than some 
critical value ($\delta_{cr}^{\rm (SGW)}=1.5$) (see
Fig. 7).  
This is understood by the fact that the $l$=0 mode is dominant 
for the SGW. 
Beyond the critical value, the above cancellation effect also 
reduces the energy of the SGWs.   At the critical deviation, the time
lag  between the
longest and shortest  radii  is almost the same as the most dominant 
QNM period. 
If the deviation is larger than the critical value, i.e., if the time
lag is  larger than the most dominant QNM period, 
 the total energy is reduced by a phase cancellation.
In fact, the difference between the values of $\delta_{cr}^{\rm (TGW)}$ and $\delta_{cr}^{\rm (SGW)}$ is explained by the difference between QNM periods of TGWs and SGWs.

For the case of a rotating black hole, we show the results in 
Fig. 8.  
The results are similar to those of the Schwarzschild case except for 
the following two points. First, since the system is  not completely spherically symmetric, 
the TGW is
also generated for any values of $\delta$. There exists some value of
$\delta_0$, which is not zero and depends on the rotation parameter $a$ 
(e.g. 
$\delta_0 \sim 0.02$ for $a=0.5$, 
$\delta_0 \sim 0.08$ for $a=0.9$, and
$\delta_0 \sim 0.10$ for $a=0.99$), at which the emitted energy of TGWs 
takes a
minimum value.   As $a$ increases, its minimum energy  and  the 
corresponding deformation parameter $\delta_0$ increase 
(see Fig. 8(b)).  
The increase of  minimum energy is easily understood 
because of a large deviation from spherical
symmetry.  The fact that $\delta_0$ increases as $a$ increases may 
also be 
explained by the following reason.  When we estimate the falling time
of a test particle from a finite radius in a Kerr spacetime, the time along the rotation axis is longer than
that on the equatorial plane if the initial radii are equal.  
Then if the dust shell is slightly oblate, the time lag between test particles
along the rotation  axis and
on the equatorial plane becomes smaller, which is close to a spherically symmetric case,   and then a phase cancellation
effect will reduce
the emitted gravitational radiation. As for the SGW in a Kerr geometry, the
results are almost same as those in a Schwarzschild one. 

Secondly, the existence of a non-vanishing minimum value of TGW is important for
observation.  Since the SGW does not depend on $\delta$, we can find
how  much
deviation is allowed for SGW to dominate TGW from Figs. 7 
and 8. That is,  the energy of
the SGW is  larger than that of the TGW if $\delta_1 < \delta < \delta_2$ and $\omega_{BD} =500$
($\delta_1=-0.07,~\delta_2=0.07$ for $a=0$, 
$\delta_1=-0.05,~\delta_2=0.09$ for $a=0.5$,
$\delta_1=0.03,~\delta_2=0.11$ for $a=0.9$, and 
$\delta_1=0.05,~\delta_2=0.12$ for $a=0.99$ when 
$\omega_{BD}=500$).   The
emitted SGW is proportional to $1/\omega_{BD}$ if $\omega_{BD} \gg
1$.    Then if $\omega_{BD}$ is larger than some critical value 
$\omega_{BD}^{\rm cr}$, which
depends on $a$, the SGWs never dominate TGWs 
($\omega_{BD}^{\rm cr} =20,000$ for $a=0.5$, 
$\omega_{BD}^{\rm cr} =1,250$ for $a=0.9$, and 
$\omega_{BD}^{\rm cr} =750$ for $a=0.99$).  We may speculate
that in  generic
gravitational collapse the SGW does not dominate the TGWs if
$\omega_{BD}>$ several thousand.

\section{Concluding Remarks}

   In this paper, analyzing test particles around a Kerr black hole, 
we studied both scalar and tensor modes of gravitational waves in
BD theory.  We examined two models: A test particle plunging
with a constant azimuthal angle into a Kerr black hole, and a
non-spherical dust shell collapsing into a Kerr black hole.
In the first model, we show that the SGWs have little dependence 
on the the rotation parameter $a$ and the azimuthal angle $\theta$ 
because the $s$-wave is dominant.  
If a coupling constant of BD theory $\omega_{BD} \gsim 500$, 
however, the
TGWs dominate the SGWs except for one very limited case, that is, an
observation in the face-on direction of a test particle falling into a
Schwarzschild black hole.
In the second model, we also show that the SGWs dominate only for
the highly spherically symmetric case.  In particular, for the case of a
rotating black hole, the emitted energy  for TGWs takes 
a non-zero minimum value  when the deformation is slightly oblate. 
Since  that for SGW is less sensitive to a deviation from a spherical 
symmetry,
the SGW cannot be dominant unless
$\omega_{BD} \lsim 750 $ for 
$a/M=0.99$ or unless  $\omega_{BD} \lsim 20,000 $ for 
$a/M=0.5$.  

Here, we remark on the detectability of SGWs. 
In the First/Advanced LIGO operation, the observation limit of the 
gravitational wave amplitude is expected to be
$h>10^{-21}$ for the $50\sim200$ Hz band, and 
$h>10^{-23}$ for the $10\sim1,000$ Hz band, respectively \cite{LIGO}.

First we mention that the QNM frequency provides 
some naive measure for the observable range.
 The relationship between 
the observed  frequency 
$f$ and the  QNM frequency $\omega$ is 
\begin{eqnarray}
f & \simeq & 3.30 \times 10^{3} \left( \frac{M_{\odot}}{M} 
\right) M\omega ~~~ [{\rm Hz}].
\end{eqnarray}
for typical   observation of
 NS  ($M \simeq 1.4M_{\odot}$)
 and BH ($M \sim 10 M_{\odot}$). The most dominant 
contribution of QNMs is $n=1$, for which we find ${\rm Re}~M\omega
\sim  0.37$
for TGW and ${\rm Re}~M\omega \sim 0.11$ for SGW.
Hence, for the BH-NS system, the observable frequencies in First
(Advanced)  LIGO are
$M\omega=0.02 \sim 0.08$ ($0.004 \sim 0.4$), while for the BH-BH
system, 
$0.15 \sim 0.61$($0.03 \sim 3.04$). Both 
observations of SGW and TGW are within preferable ranges of frequency.

Secondly,  we estimate the maximum amplitude 
of gravitational  radiation. The expected amplitudes of SGW and TGWs
 ($h^{\rm (TGW)}, h^{\rm (SGW)}$)
are
\begin{eqnarray*} h^{\rm (TGW)} & \simeq & 2 \times 10^{-20} \left( 
\frac{\mu}{M} 
\right) 
\left(\frac{1{\rm Mpc}}{r} \right) ,\\
h^{\rm (SGW)} & \simeq & 1 \times 10^{-22} \left( 
\frac{500}{\omega_{BD}} \right)
\left( \frac{\mu}{M} \right) \left( \frac{1{\rm Mpc}}{r} \right) ,
\end{eqnarray*}
where we set $h_{+}=0.4$ and $\phi^{(1)}=0.002$ from our results.

Fixing the BD coupling constant $\omega_{BD} = 500$,
we find that  the
limiting distance of SGWs  in First LIGO for  NS-NS, BH-BH collision
is 
 $r$=100kpc, while $r$=20Mpc for TGWs.
This limiting  distance becomes much longer when we have the  NS-BH
system  and  
Advanced LIGO detector, i.e., the target for SGW could set farther than the 
Virgo  Cluster (20Mpc).

Conversely, 
if the SGW is observed in a gravitational wave 
detector with polarization information,  what kind of constraint on  the BD
coupling constant
$\omega_{BD}$ we will find ? We can at least discuss the upper bound.
 For example, we can
easily  say that if NS-BH  collision
  occurred in our Galaxy and 
the SGW were observed, we could determine definitely the  BD
coupling constant even for much higher values than 500 (see Table 1).

Although our analysis has been done by test particles, we believe that
our conclusions for the SGWs are valid for more generic cases. As for other
scalar-tensor theories of gravity, we expect similar results, although we  may need further analysis \cite{HCNN}.
 In order to determine a theory of gravity by gravitational wave
detection, the above results require observations of  polarizations and 
extractions of
information of the scalar mode from observed data, for which we need several
observational sites in the world. 

It is very interesting to test  theories of gravity by  direct
observation  of gravitational radiation.
It will open a new window in gravitational astronomy, which may
provide a third and the final observational eye to see the Universe and
may be  expected  to reveal
some problems in fundamental physics (such as the equation of state at
high  density) as well as to determine many properties and orbital
elements  of compact objects.

\acknowledgments

  We would like to thank Takashi Nakamura, Masaru Shibata and Hideyuki
Tagoshi for useful comments. We also would like to thank Paul Haines
for critical reading of our manuscript. 
This work was supported partially by the
Grant-in-Aid for Scientific Research  Fund of the
Ministry of Education, Science and Culture  (Specially Promoted
Research No. 08102010), and by a Waseda University
Grant for Special Research Projects.

\newpage
\appendix
\section{Source Term of Gravitational Waves by a Test Particle in
a Kerr Spacetime}

 In \S. 2, we summarize the tensor gravitational waves
from a test particle in a Kerr spacetime.  In order to solve the
basic equation (the Sasaki-Nakamura equation), we first have to give
the  source
term $S_{lm\omega}$ in Eq. (\ref{eqn:GeneRW}).  Nakamura and Sasaki
gave  the
explicit form for a test particle moving on  
the rotation axis \cite{SN} and Kojima and Nakamura gave it on the equatorial plane \cite{KN}. 
Shibata presented it for a motion with
$r$ = constant \cite{Shib}. 
 However, no one has so far written  it down for 
generic
motions.  Since we had to calculate the case with off-plane
motion ($\theta$ = constant), here we present the explicit form
of a source term for generic motions in use of numerical calculation.  Here 
we consider only
non-periodic motion or unbound system.  If the system is
bounded, we may find many turning points, for which our
expression is no longer valid.  We have to rewrite the orbital
motion in terms of time (either a coordinate time $t$ or a proper
time $\tau$) instead of $r$. This may be straightforward
\cite{footnote3}.

A test particle in a Kerr spacetime is described by the equations of
motion as 
\begin{eqnarray}
\Sigma{dr \over d\tau}&=&\pm\sqrt{R},\\
\Sigma{d\theta \over d\tau}&=&\pm\sqrt{\Theta},\\
\Sigma{d\varphi \over d\tau}&=&-\left( a E -
\frac{L_{z}}{\sin^{2}\theta} 
\right) + \frac{a}{\Delta}P,\\
\Sigma{dt \over d\tau}&=&-a (a E \sin^{2}\theta - L_{z}) + 
\frac{r^{2}+a^{2}}{\Delta}P,
\end{eqnarray}
where
\begin{eqnarray}
P &=& E (r^{2}+a^{2}) - a L_{z},\\
R &=& P^{2} - \Delta \left( r^{2}+(L_{z}-a E)^{2} + C\right),\\
\Theta &=& C - \cos^{2}\theta \left( a^{2} (\mu^2-E^{2}) +
\frac{L_{z}^{2}}{
\sin^{2}\theta} \right),
\end{eqnarray}
and 
$E$ and $L_z$ are conserved energy and $z$-component angular
momentum of a test particle, respectively,  and $C$ is the Carter constant. We
can  easily see that $\theta$ =
constant is a geodesic if
$E=\mu$ and $L_z=0$.

The source term 
$S_{lm\omega}$ is given by 
\begin{eqnarray}
S_{lm\omega} & = & \frac{\gamma\Delta}{(r^{2}+a^{2})^{3/2}r^{2}} {\cal W} 
\exp (-i\int^{r}\frac{K}{\Delta} dr) .
\end{eqnarray}
Here ${\cal W}$ is divided into three parts as 
\begin{eqnarray}
{\cal W} & = & {\cal W}_{nn}+{\cal W}_{\bar{m}n}
+{\cal W}_{\bar{m}\bar{m}},\\ 
{\cal W}_{nn} & = &
f_{0}(r) e^{i\chi (r)}+\int^{\infty}_{r}dr' f_{1}(r') e^{i\chi (r')}+
\int^{\infty}_{r}dr'\!\!\!\int^{\infty}_{r'}dr'' f_{2}(r'') e^{i\chi 
(r'')},
\\
 {\cal W}_{\bar{m}n} & = &
g_{0} (r) e^{i\chi (r)}+\int^{\infty}_{r}dr' g_{1}(r') e^{i\chi (r')}+
\int^{\infty}_{r}dr'\!\!\!\int^{\infty}_{r'}dr'' g_{2}(r'') e^{i\chi 
(r'')},
\\
 {\cal W}_{\bar{m}\bar{m}} & = & h_{0}(r) e^{i\chi (r)}+
\int^{\infty}_{r}dr' h_{1}(r') e^{i\chi (r')}+
\int^{\infty}_{r}dr'\!\!\!\int^{\infty}_{r'}dr'' h_{2}(r'') e^{i\chi 
(r'')},
\end{eqnarray}
where
\begin{eqnarray}
\chi & = & \omega(t+r^{*})-m\tilde{\varphi} ,\\
 f_{0} &
= & - \frac{1}{\omega^{2}} w_{nn} ,\\ 
f_{1} & = & -
\frac{2}{\omega^{2}} \left( \frac{dw_{nn}}{dr}  + 
i(2a\omega\sin^{2}\theta-m)\frac{d\tilde{\varphi}}{dr} w_{nn} \right)
,\\ 
f_{2} & = & - \frac{1}{\omega^{2}} \left[
\left(\frac{d^{2}w_{nn}}{dr^{2}}
 - 2im\frac{d\tilde{\varphi}}{dr}\frac{dw_{nn}}{dr} -
im\frac{d^{2}\tilde{\varphi}}{dr^{2}}w_{nn}-
m^{2}\left(\frac{d\tilde{\varphi}}{dr}\right)^{2}w_{nn} \right)
\right. 
\nonumber \\
 & &
 +2ia\omega\sin^{2}\theta\left(\frac{dw_{nn}}{dr} 
\frac{d\tilde{\varphi}}{dr} + w_{nn}
\frac{d^{2}\tilde{\varphi}}{dr^{2}}-
im\left(\frac{d\tilde{\varphi}}{dr}\right)^{2} w_{nn} \right)
\nonumber \\
 & &\left. -a^{2}\omega^{2}\sin^{4}\theta
\left(\frac{d\tilde{\varphi}}{dr}\right)^{2} w_{nn}\right] ,\\
 g_{0} & = &
\frac{i}{\omega} \rho~w_{\bar{m}n}^{(1)}~w_{\bar{m}n}^{(2)} ,\\ 
g_{1} & = & \frac{i}{\omega} w_{\bar{m}n}^{(1)}
\left[ 
-w_{\bar{m}n}^{(3)} + w_{\bar{m}n}^{(2)} \frac{d\rho}{dr} 
+2\rho\frac{dw_{\bar{m}n}^{(2)}}{dr} -i
\left(m-a\omega \sin^{2}\theta\right)\rho~w_{\bar{m}n}^{(2)}
\frac{d\tilde{\varphi}}{dr} \right] ,\\
g_{2} & = & -\frac{i}{\omega} w_{\bar{m}n}^{(1)} \left[  
\frac{d}{dr}\left(w_{\bar{m}n}^{(3)}  + \rho \frac{dw_{
\bar{m}n}^{(2)}}{dr} 
\right) \right.\nonumber \\
&&\left. -i \left(m-a\omega\sin^{2}\theta\right)
\left(w_{\bar{m}n}^{(3)}-\rho
\frac{dw_{\bar{m}n}^{(2)}}{dr}\right)
\frac{d\tilde{\varphi}}{dr} \right] ,\\
h_{0} & = & -\frac{1}{2}
\frac{r^{2}\bar{\rho}^{2}}{|dr/d\tau|}\mbox{}_{- 2}S_{lm}^{a\omega} 
(w_{\bar{m}n}^{(1)})^2
,\\
h_{1} & = & -\frac{1}{2}
\frac{\rho\bar{\rho}^{2}}{|dr/d\tau|}\mbox{}_{- 2}S_{lm}^{a\omega} 
(w_{\bar{m}n}^{(1)})^2
\left( \frac{d}{dr}\left(\frac{r^{2}}{\rho}\right) +
\rho^{-4}\frac{d}{dr}\left(\rho^{3}r^{2}\right) 
\right),\\
h_{2} & = & -\frac{1}{2}
\frac{\rho\bar{\rho}^{2}}{|dr/d\tau|}\mbox{}_{- 2}S_{lm}^{a\omega} 
(w_{\bar{m}n}^{(1)})^2
\frac{d}{dr}\left(\rho^{-4}\frac{d}{dr}(\rho^{3}r^{2})\right) 
\end{eqnarray}
with
\begin{eqnarray}
\tilde{\varphi} & = & \varphi + \int^{r} \frac{a}{\Delta}dr ,\\
\rho & = & \frac{1}{r-ia\cos\theta},\\
w_{nn} & = & -\frac{\mu r^{2}}{2\rho \Sigma^{2}}
\left|\frac{dr}{d\tau}
\right| 
{\cal L}^{\dag}_{1} [\rho^{-4}{\cal
L}^{\dag}_{2}(\mbox{}_{-2}S_{lm}^{a\omega}\rho^{3})] ,\\
 w_{\bar{m}n}^{(1)} & = &
\sqrt{\Theta} + i\sin\theta (aE-\frac{L_{z}}{\sin^{2}\theta}), \\
w_{\bar{m}n}^{(2)} & = & \frac{r^{2}\bar{\rho}}{\rho^{2}}{\cal
L}^{\dag}_{2}
(\rho\bar{\rho}\mbox{}_{-2}S_{lm}^{a\omega}),\\ 
w_{\bar{m}n}^{(3)} &
= & \frac{1}{2}r^{2}\rho {\cal L}^{\dag}_{2} 
\left(\rho^{3}\mbox{}_{-2}S_{lm}^{a\omega}
\frac{d}{dr}(\bar{\rho}^{2}\rho^{-4})\right),
\end{eqnarray}
where an overbar denotes the complex conjugate.
The operator 
${\cal L}^{\dag}_{s}$ is defined by 
\begin{equation}
 {\cal L}^{\dag}_{s} = 
\frac{\partial}{\partial\theta} +  a\omega\sin\theta - 
\frac{m}{\sin\theta} + s\cot\theta .
\end{equation}

\newpage
\vspace{0.5cm}
\begin{flushleft}
{\bf Figure Captions}
\end{flushleft}
\baselineskip .65cm
\vskip 0.1cm

\begin{figcaption}{fig:ins1}{12cm}
Waveforms of TGWs by a  test particle with constant azimuthal angle $\theta$ 
falling into a Kerr black hole.
We show two modes ($h_+$ and $h_\times$) of TGWs for $\theta=\pi/6$ ((a), (b)), =
 $\pi/3$ ((c), (d)) and  $h_+$ mode for $\theta=\pi/2$ ((e))
($h_\times$ vanishes.).  We choose a Kerr parameter $a$= 0 (solid line), 0.5 (dashed line), 0.9 (dash-dotted line) and 0.99 (dotted line).
\label{fig1}
\end{figcaption}
\begin{figcaption}{fig:ins2}{12cm}
Emitted energy of TGWs showen in Fig. 1.\\
Circle (filled), square, triangle and circle (open) represents the cases of $a=0$, $a=0.5$, $a=0.9$ and $a=0.99$, respectively.
\label{fig2}
\end{figcaption}
\begin{figcaption}{fig:ins3}{12cm}
Waveforms of SGWs by a  test particle with constant azimuthal angle $\theta$ 
 falling into a Kerr black hole for $\omega_{BD}=500$. We show SGWs for
$\theta=\pi/6$ ((a)), = $\pi/3$ ((b)) and  = $\pi/2$ ((c)).  We choose a Kerr
parameter $a$= 0 (solid line), 0.5 (dashed line) and 0.9 (dash-dotted line).  
We find that they do not depend very much on the azimuthal angle and the Kerr parameter, although we find a phase difference depends on $a$.
\label{fig3}
\end{figcaption}
\begin{figcaption}{fig:ins4}{12cm}
Energy spectrum of SGWs shown in Fig. 3.\\
Solid, dashed, dash-dotted and dotted lines represent $a=0$, $a=0.5$, $a=0.9$ and $a=0.99$, respectively.
We find that the frequencies which emittes largest energy spectrum
corresponds to that of the QNM.
All lines (solid line ($a=0$), dashed one ($a=0.5$), dash-dotted one ($a=0.9$) and dotted one ($a=0.99$)) almost coincide with each other.
\label{fig4}
\end{figcaption}
\begin{figcaption}{fig:ins5}{12cm}
Emitted energy of SGWs shown in Fig. 3. \\
Circle (filled) , square, triangle and circle (open) represent 
$a=0$, $a=0.5$, $a=0.9$ and $a=0.99$, respectively.\\
 They also do not depend very much 
 on the azimuthal angle.
\label{fig5}
\end{figcaption}
\begin{figcaption}{fig:ins6}{12cm}
The tail part of the SGW is fitted by one of the QNMs, which corresponds to the
 largest peak in the energy spectrum (Fig. 4). 
Solid line represents the SGW, while
dashed line denotes the waveform fitted by one of the QNMs
((a) $a=0$, (b) $a=0.5$ and (c) $a=0.9$).
\label{fig6}
\end{figcaption}
\begin{figcaption}{fig:ins7}{12cm}
Emitted energy by a non-spherical dust shell collapsing into a Schwarzschild
 black hole ((a) prolate shell and 
(b) oblate shell) for $\omega_{BD}=500$.  A spherically symmetric shell does
not generate TGWs, while SGWs are emitted.  Since the $l=0$ mode dominates
in the SGW, the emitted energy of the SGW does not depend on
a deformation parameter $\delta$.
The SGW dominates the
TGWs for $\delta_1 (=-0.07) < \delta < \delta_2 (=0.07)$.
\label{fig7}
\end{figcaption}
\begin{figcaption}{fig:ins8}{12cm}
Emitted energy by non-spherical dust shell collapsing into a
 Kerr black hole ((a) prolate shell and (b) oblate shell) for $\omega_{BD}=500$. 
Dashed, dash-dotted and dotted lines represent $a=0.5$, $a=0.9$ and $a=0.99$ respectively.
Since the system is no longer spherically symmetric for any values of $\delta$,
 we expect the TGWs as well as the SGW.
At
$\delta_0$, which is not zero and depends on a rotation parameter $a$ 
($\delta_0 \sim 0.02,~0.08$ and 0.10 for $a=0.5,~0.9$ and 0.99, respectively),
the emitted energy of TGW  takes a minimum value.
As $a$ increases, its minimum
energy  and  the  corresponding deformation parameter $\delta_0$ increase. Since
the $l=0$ mode dominates in the SGW, the emitted energy of SGW does not depend
on  a deformation parameter $\delta$. It also depends little on $a$. 
Both energies of TGW and SGW decrease
for   $\delta \gsim \delta_{cr}^{\rm (TGW)}$ and 
$\delta \gsim \delta_{cr}^{\rm (SGW)}$ because of 
a phase cancellation effect.   The
SGW dominates the TGWs for $\delta_1 (=-0.05,~0.03$ and $0.05$ for $a=0.5,~0.9$ and $0.99) < \delta < \delta_2 (=0.09,~0.11$ and $0.12$ for $a=0.5,~0.9$ and $0.99)$.
\label{fig8}
\end{figcaption}
\newpage
\begin{table}
\begin{center}
\begin{tabular}{|c||c|c|c|}
\multicolumn{1}{|c||}{Binary System} &
\multicolumn{1}{c|}{Coma cluster} &
\multicolumn{1}{c|}{Virgo cluster} &
\multicolumn{1}{c|}{Our Galaxy} \\ 
\multicolumn{1}{|c||}{} &
\multicolumn{1}{c|}{($\sim$ 80Mpc)} &
\multicolumn{1}{c|}{($\sim$ 20Mpc)} &
\multicolumn{1}{c|}{($\sim$ 10kpc)} \\ \hline \hline
NS-NS, BH-BH &  $\omega_{BD} \lsim $  ~60 & $\lsim $ ~~250 & $\lsim $ 5 $
\times 10^5$  \\ \hline
NS-BH & $\omega_{BD} \lsim $   400 & $\lsim $ 1,700 & $\lsim $ 3.5
$\times  10^6$ \\
\end{tabular}
\end{center}
\end{table}

\begin{flushleft}
\parbox[t]{2cm}{ Table 1:\\~}\ \
\parbox[t]{14cm}
{
The upper bound of the BD coupling constant $\omega_{BD}$ if
the  SGW with a polarization information independently of the TGW
is detected.
}
\end{flushleft}


\begin{thebibliography}{99}
\bibitem{footnote1}
$G_N=1$ means, in this paper, that we set the Newtonian gravitational constant
 to be
unity. This fixes the unperturbed values of the BD scalar field as
$\phi_0= (4+2\omega_{BD})/(3+2\omega_{BD})$.  Since $\omega_{BD}>500$, 
we find that $1 \leq \phi_0 < 1.001$.
\bibitem{footnote2}
This metric perturbation $h_{\mu\nu}^{(E)}$ is the same as 
$\theta_{\mu\nu}$ in \cite{TEG}, which is used   to study 
BD theory in a weak field limit.
\bibitem{footnote3} We are now analyzing gravitational waves for
generic motions of a test particle moving off equatorial plane to study
the dependence of the eccentricity or the Carter constant of the
orbit.  It will be published elsewhere.
\end{thebibliography}
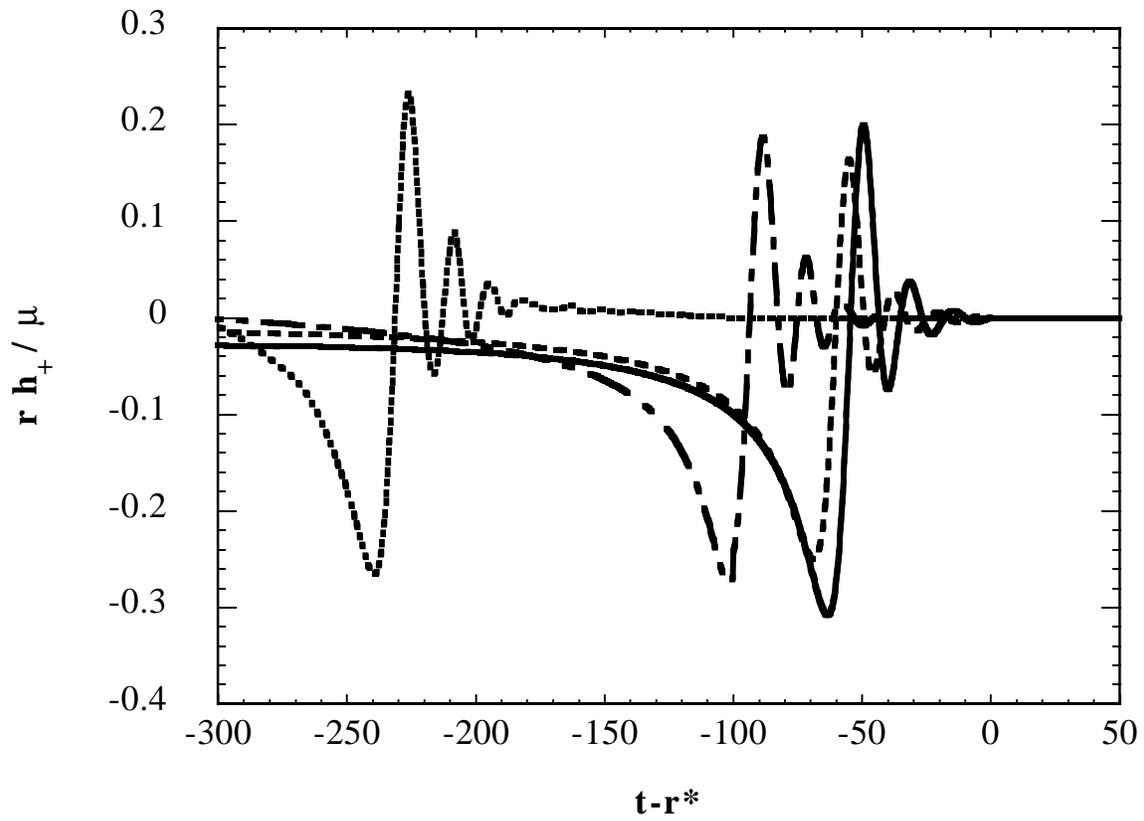
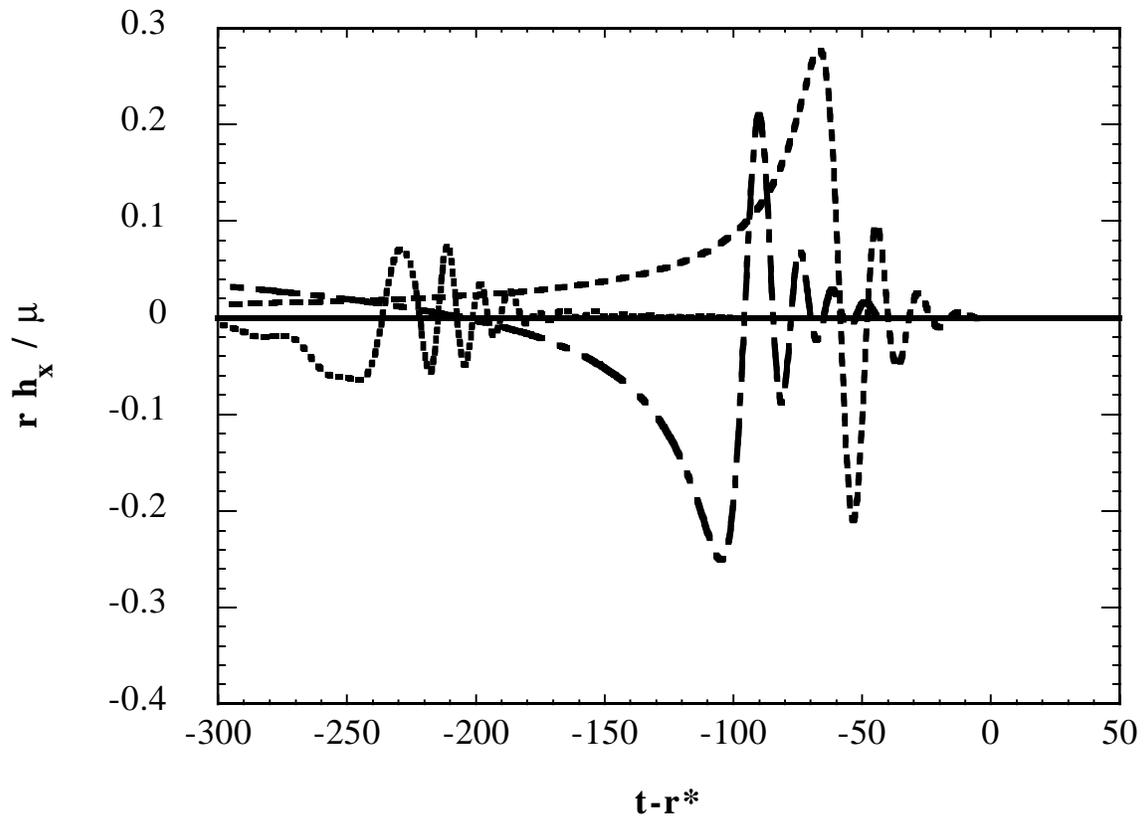
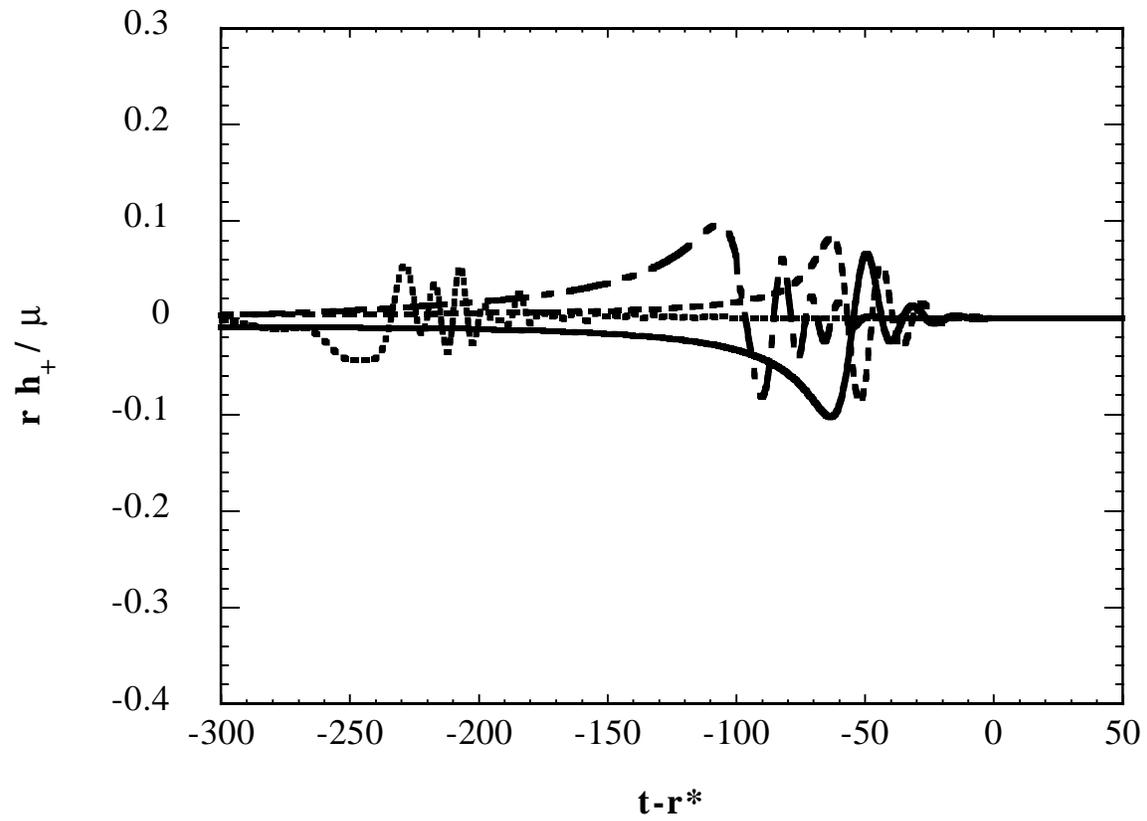
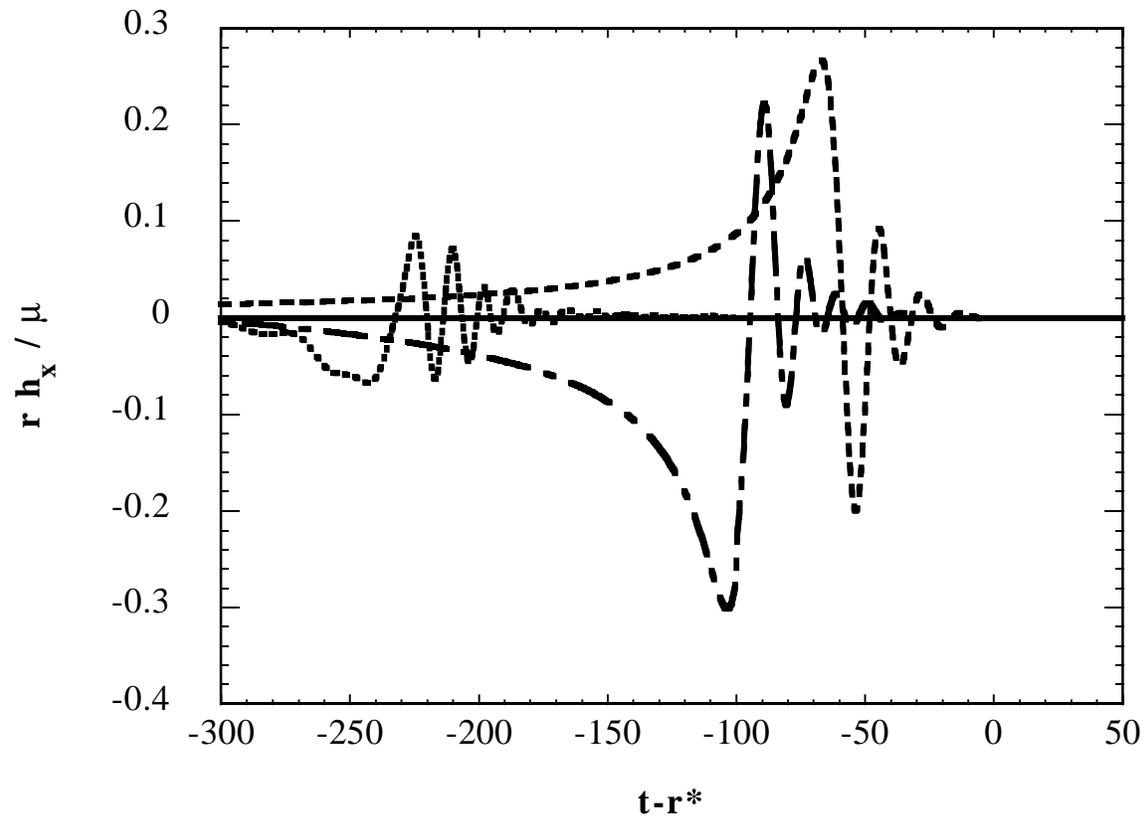
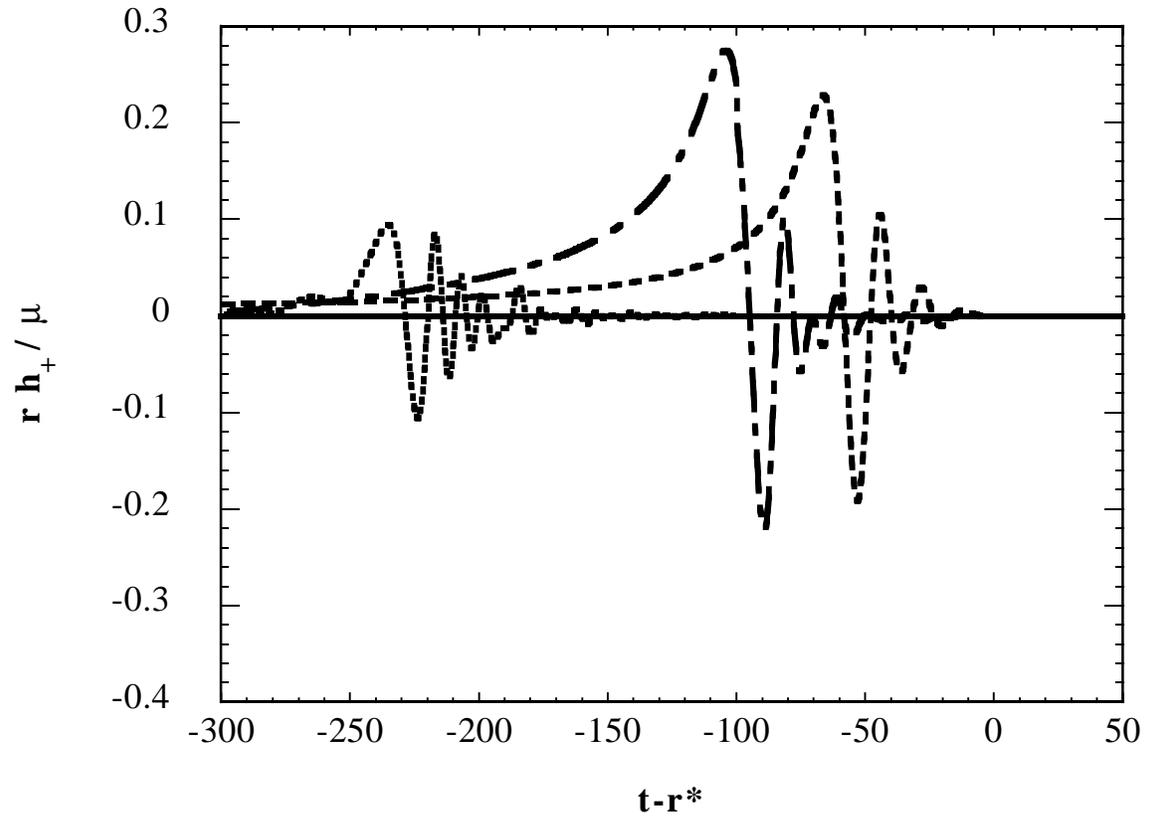
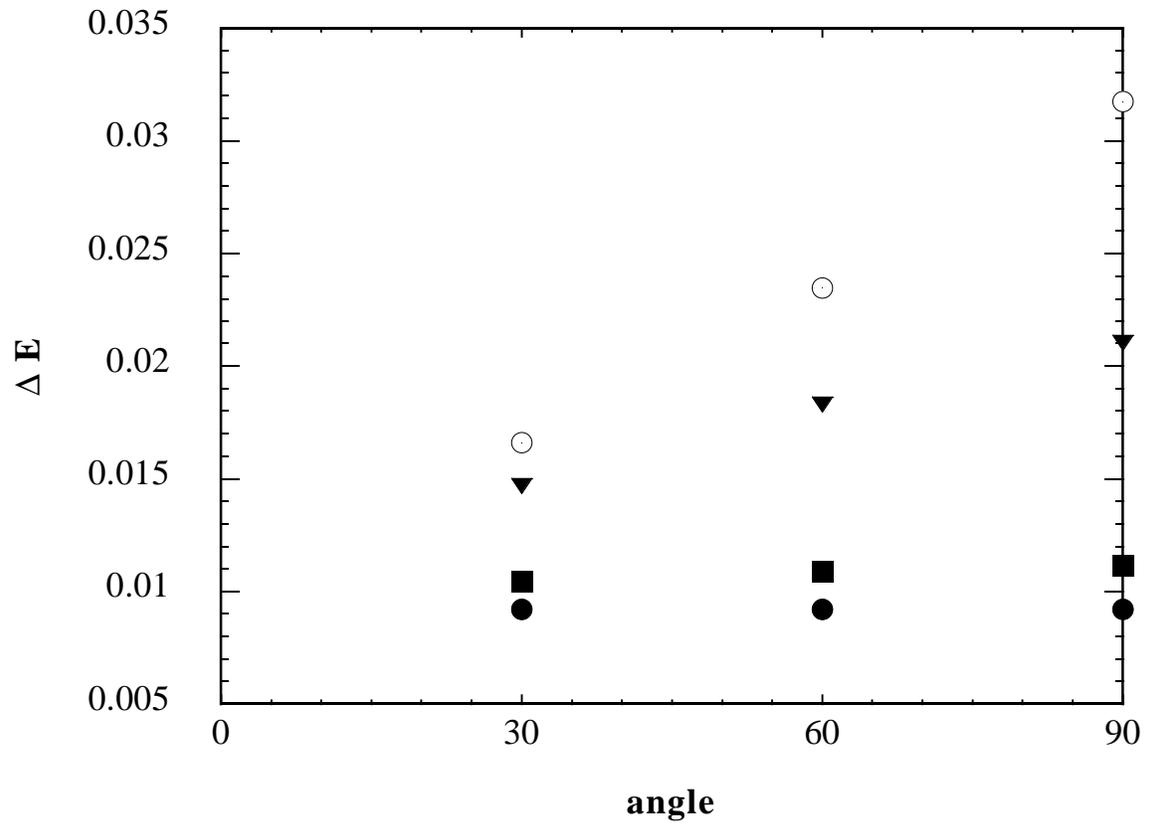
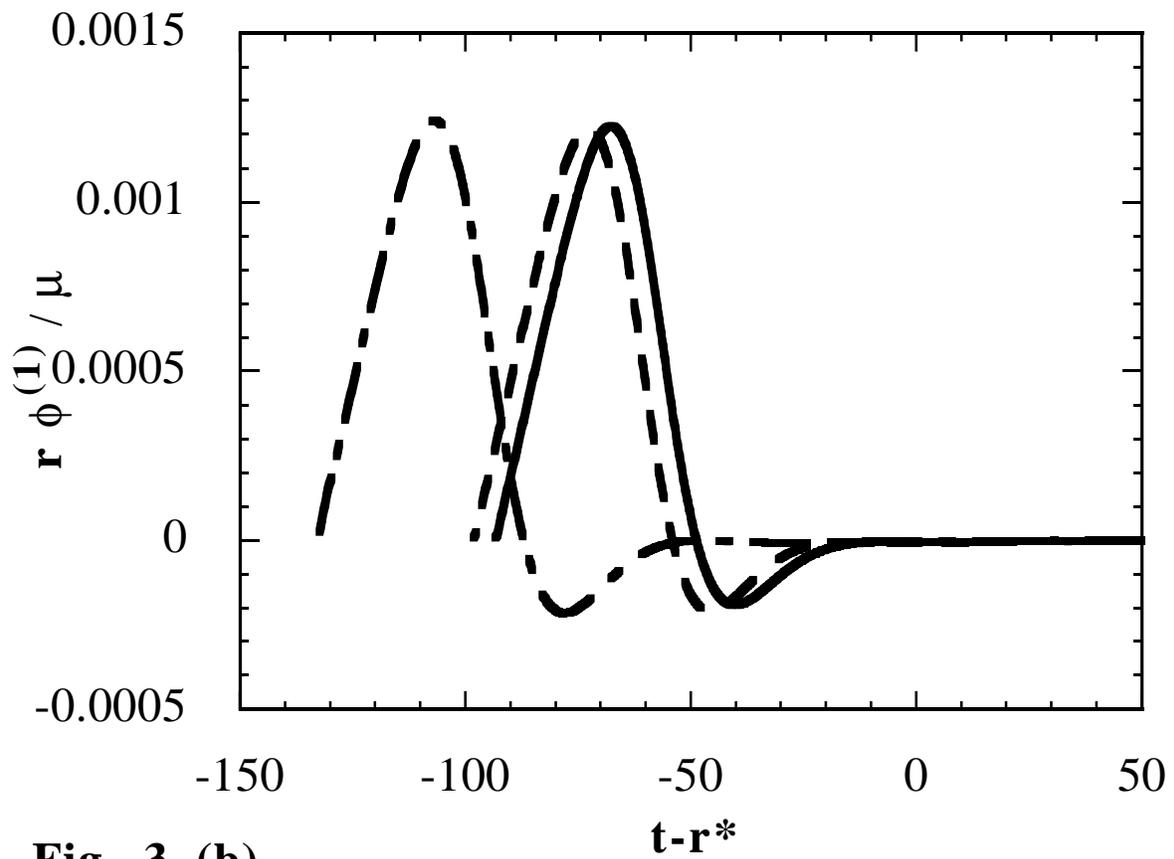
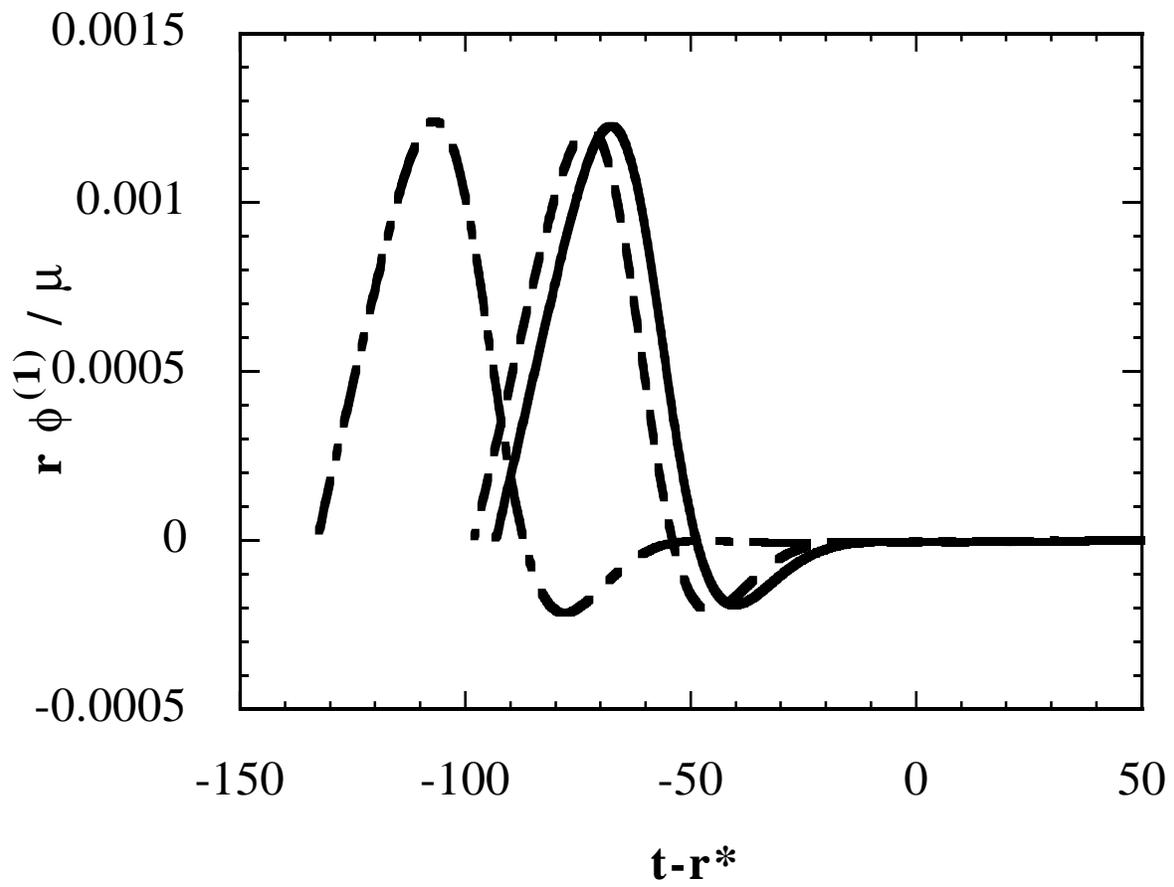
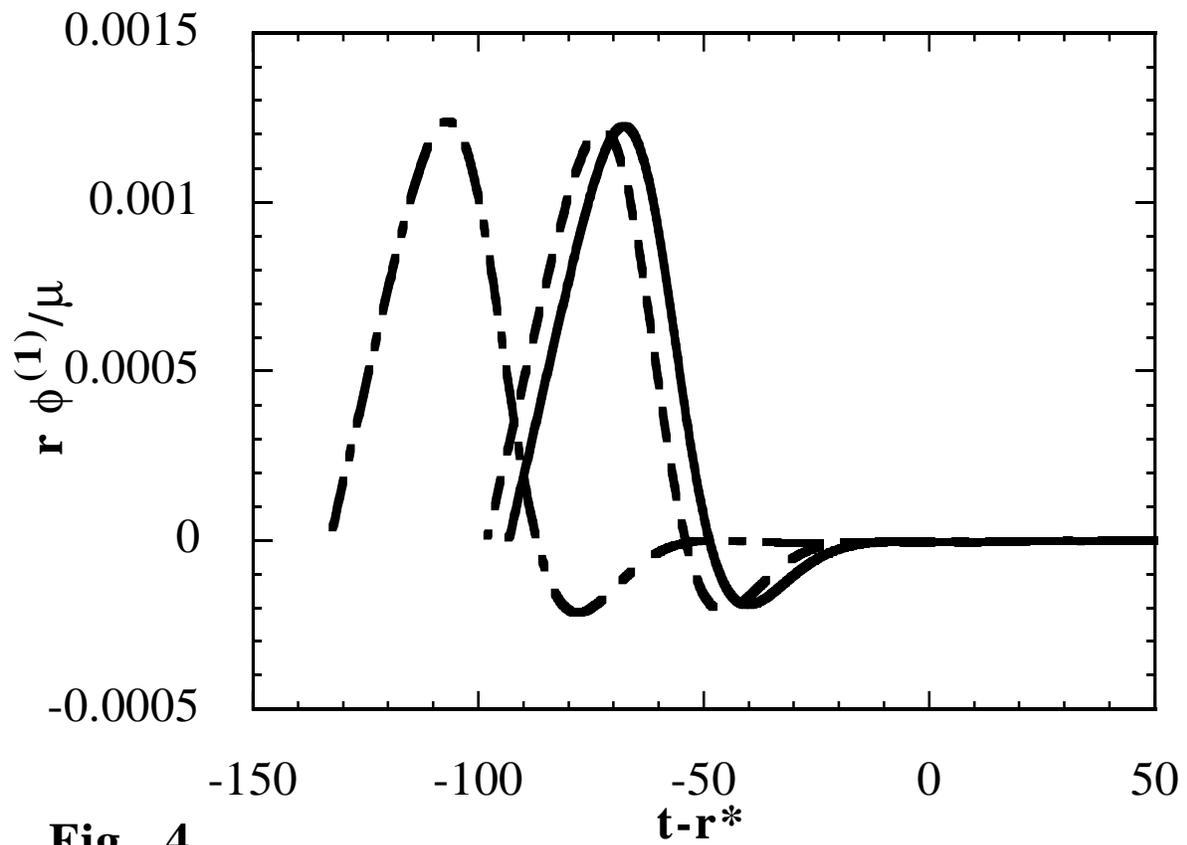
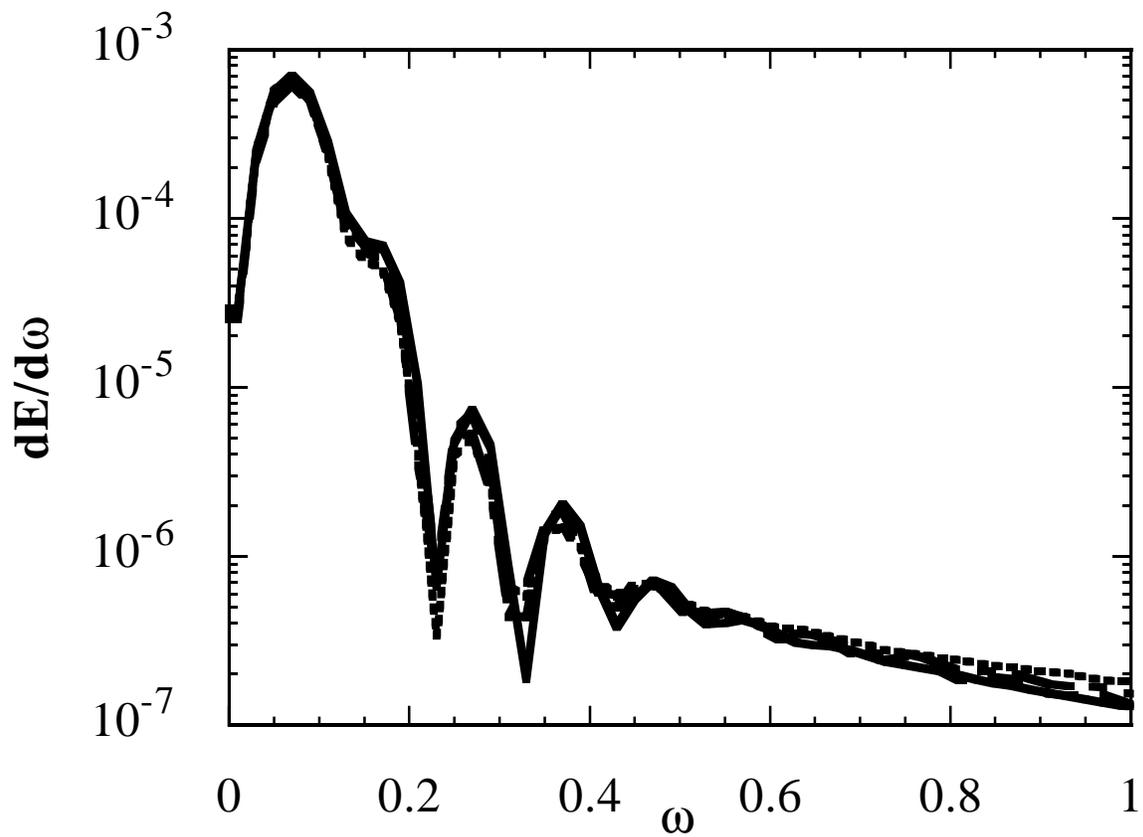
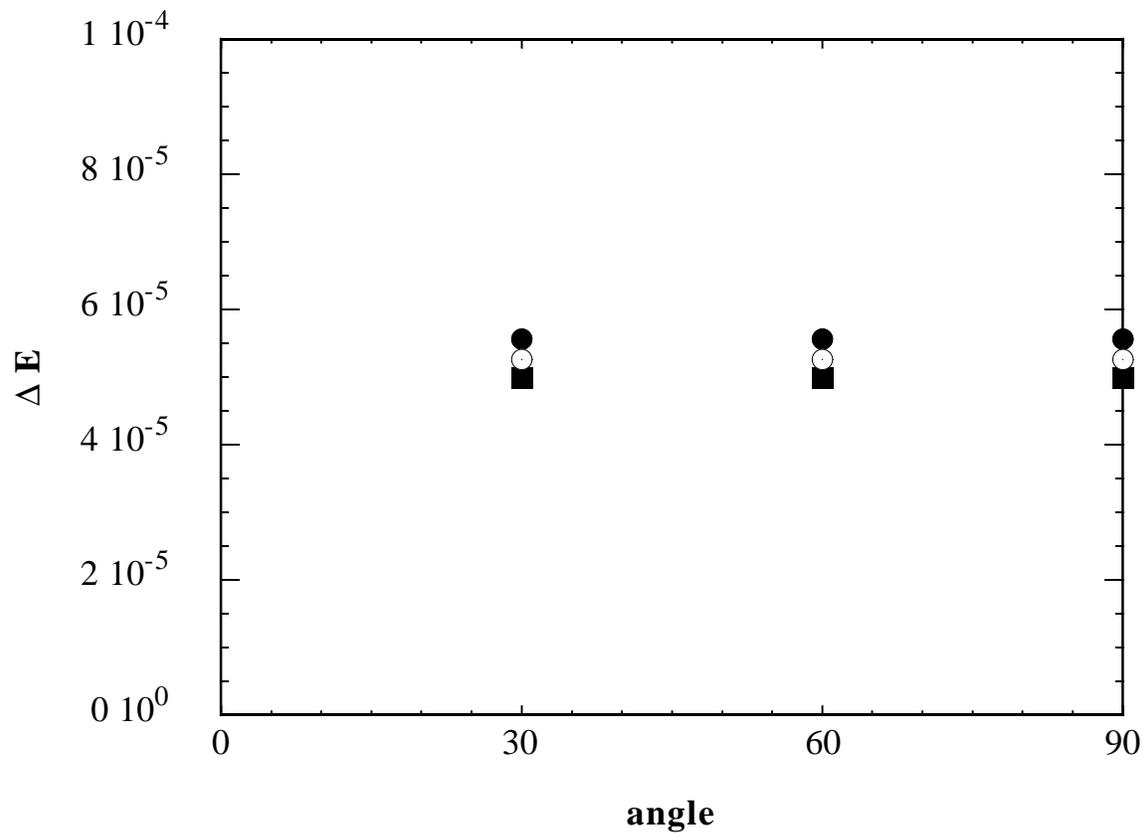
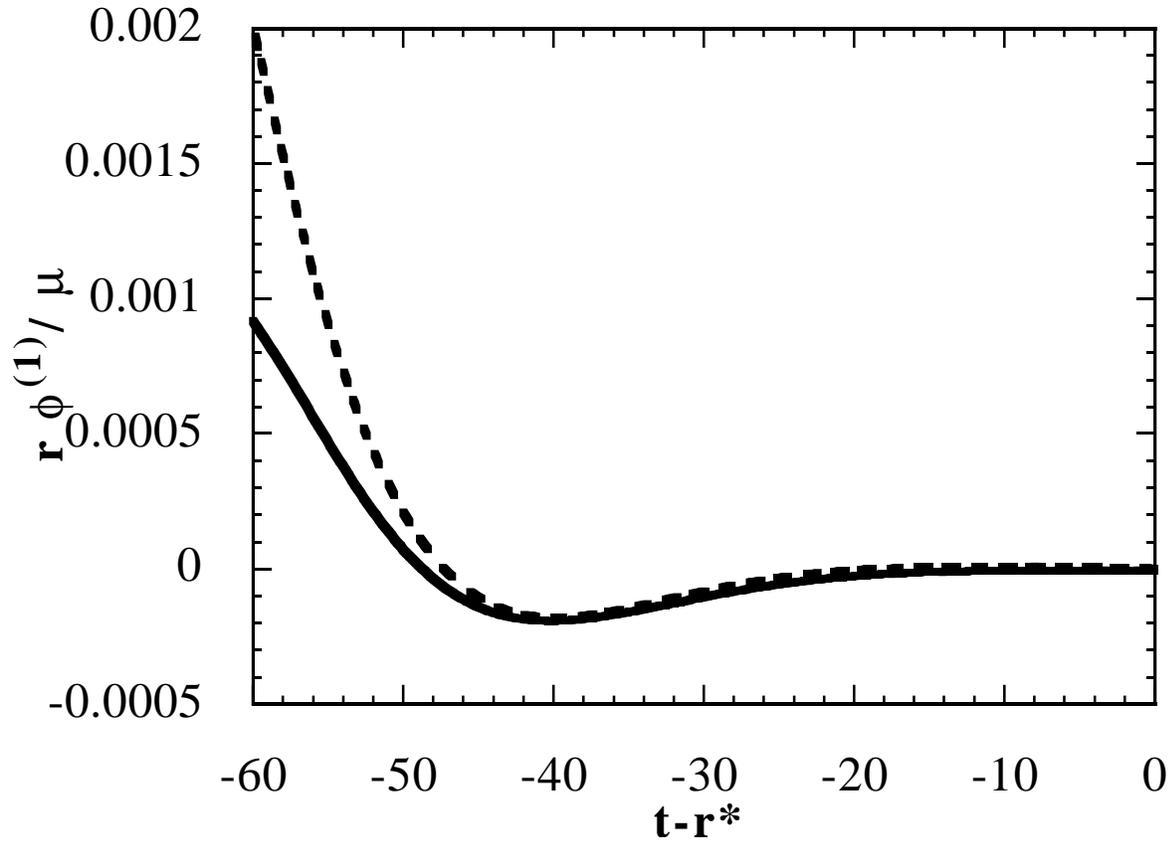
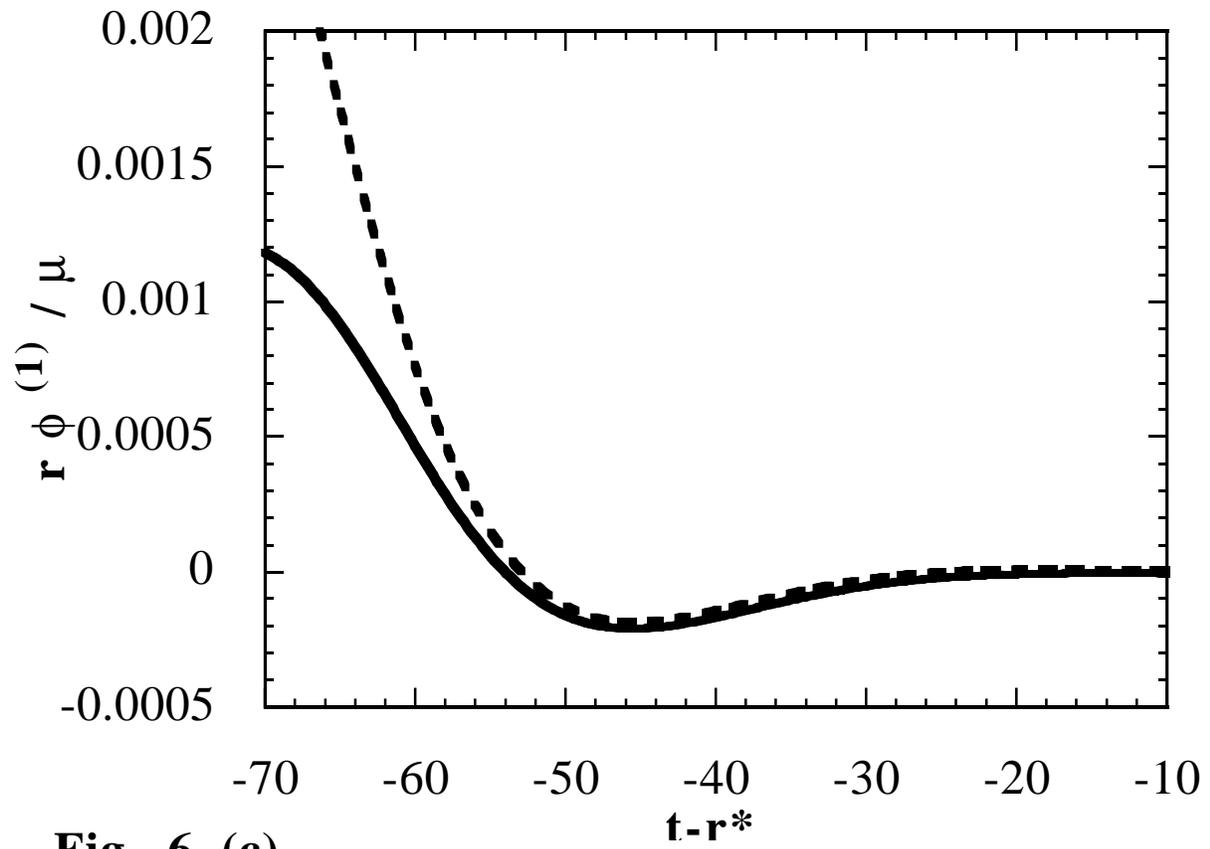
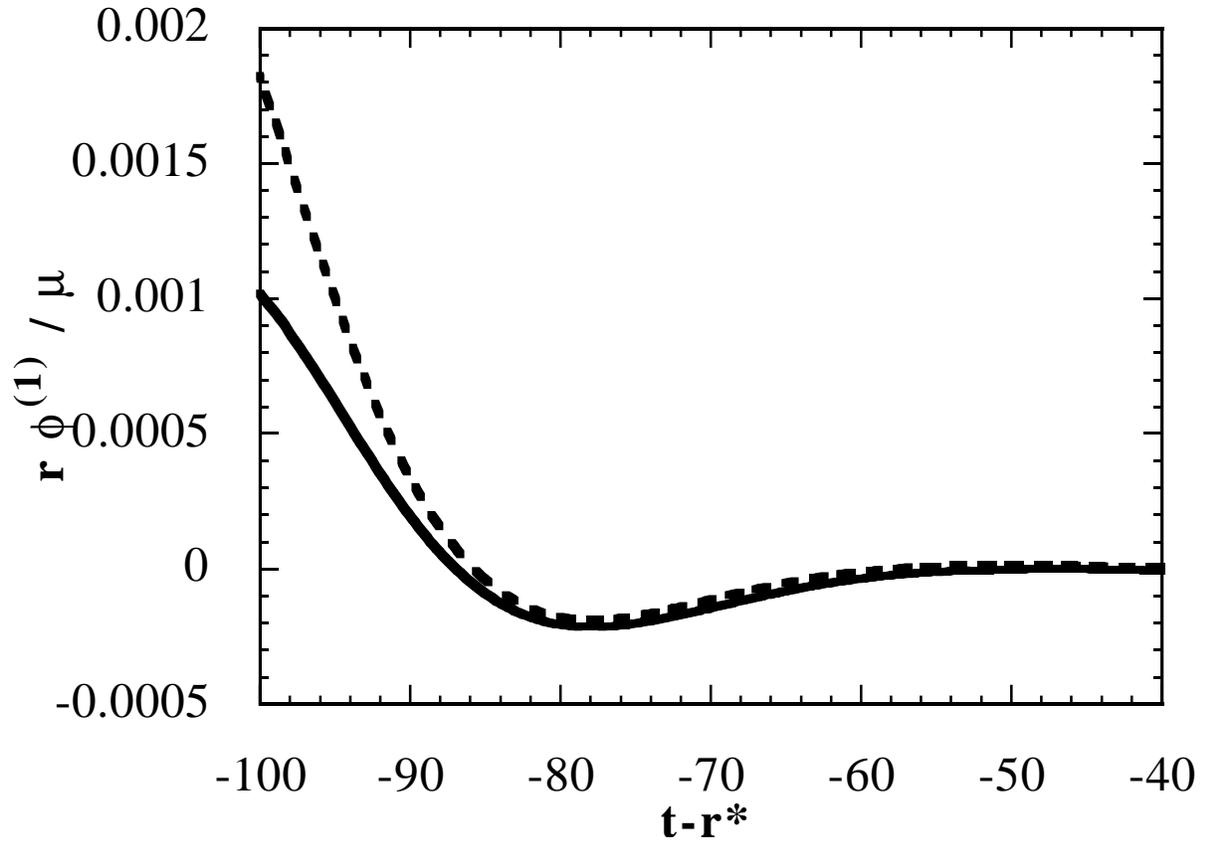
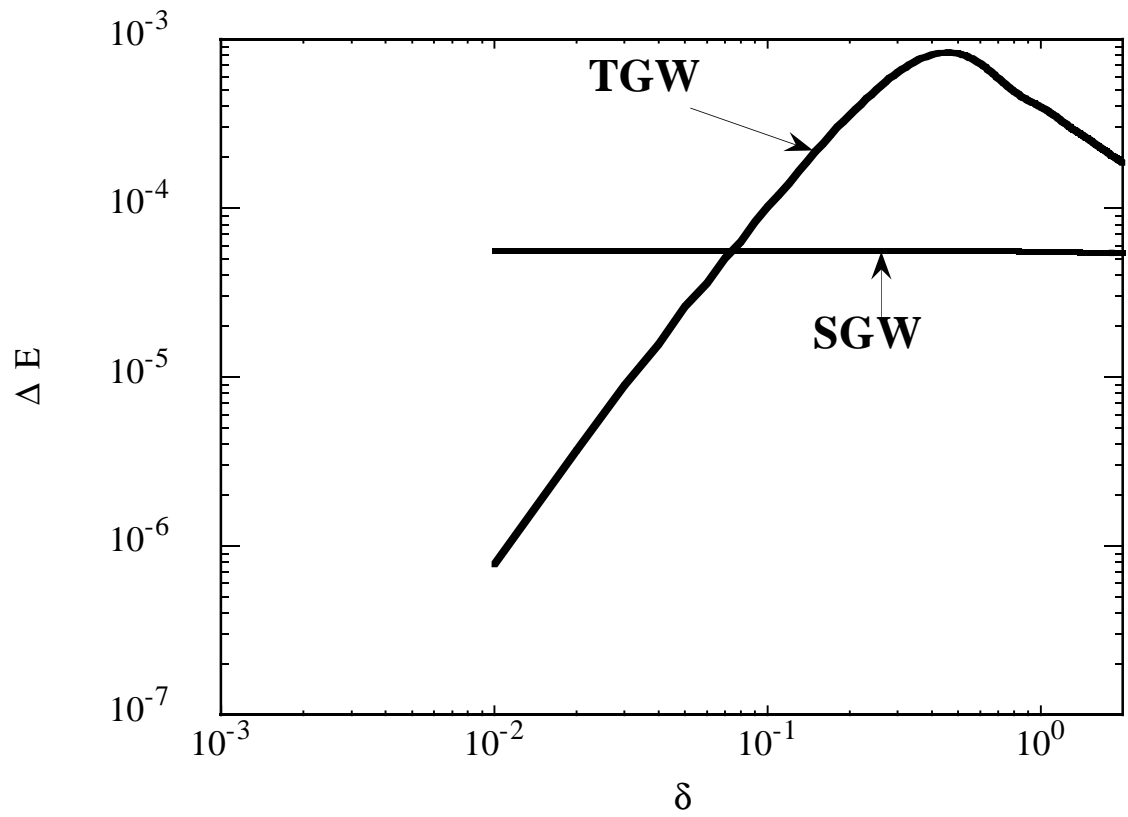
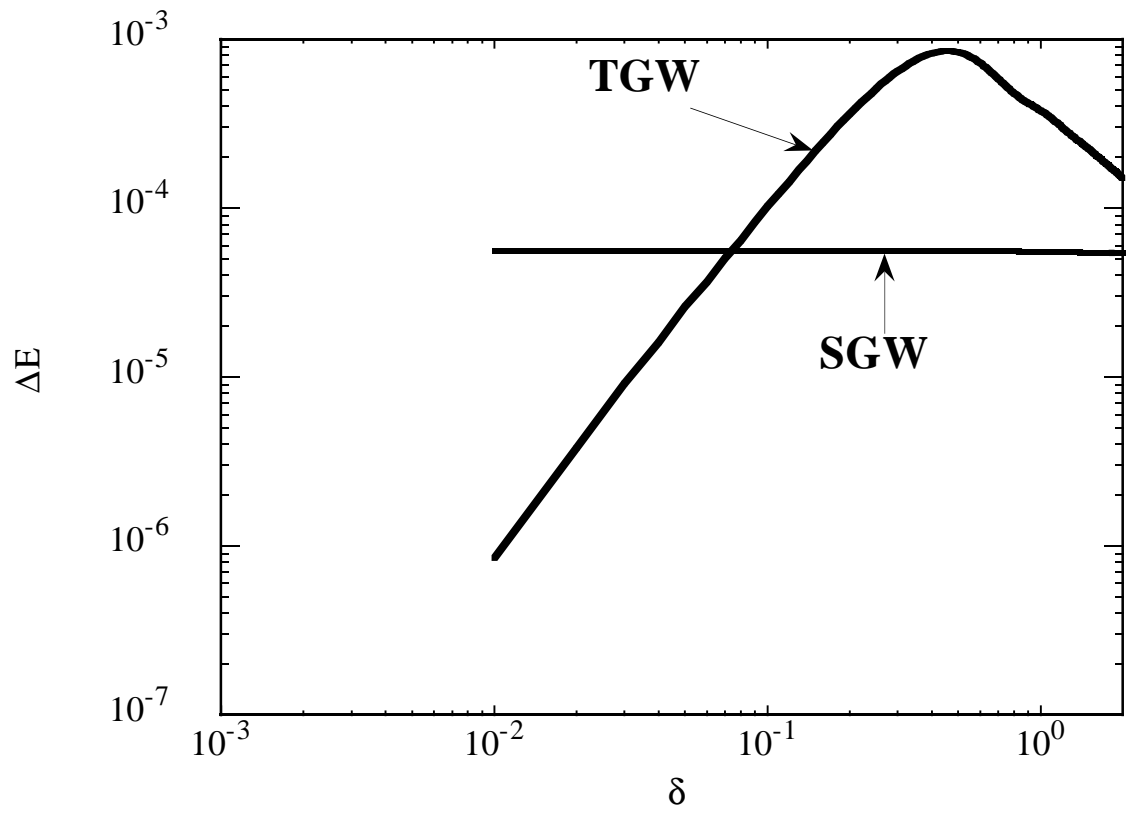
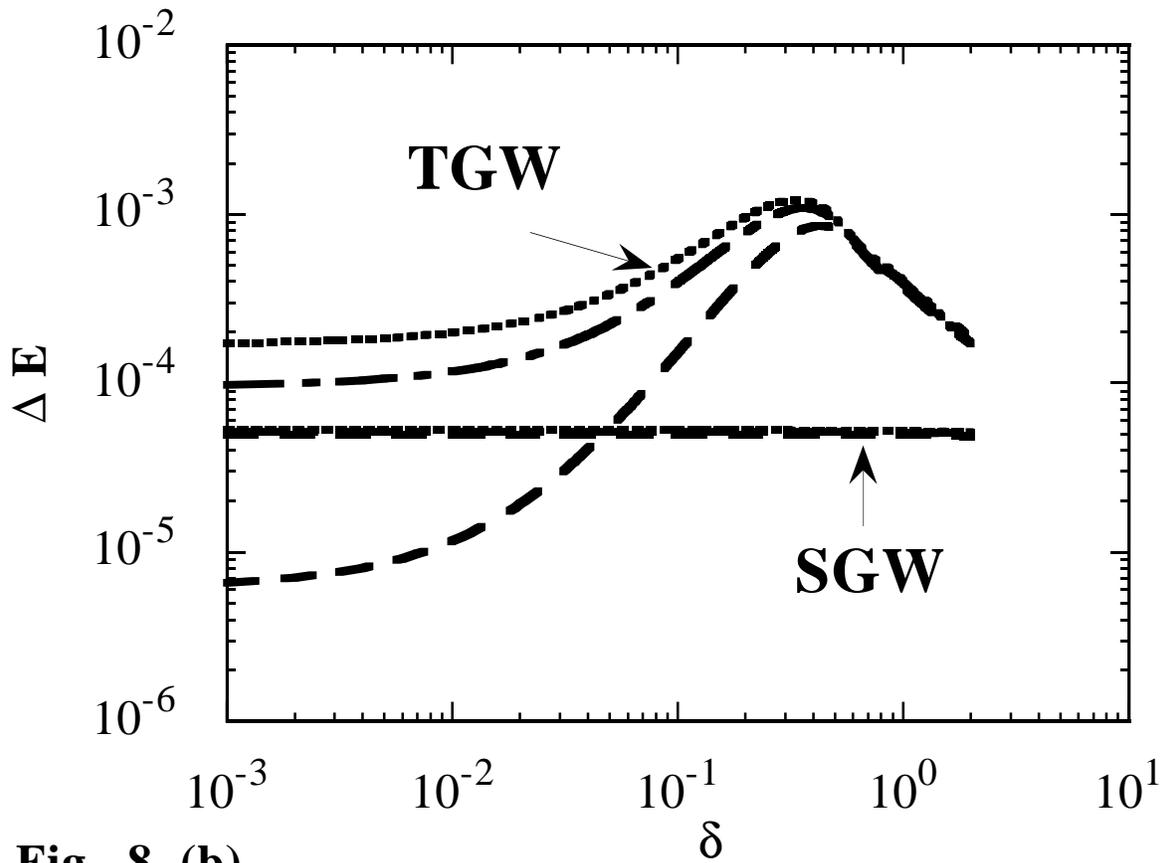
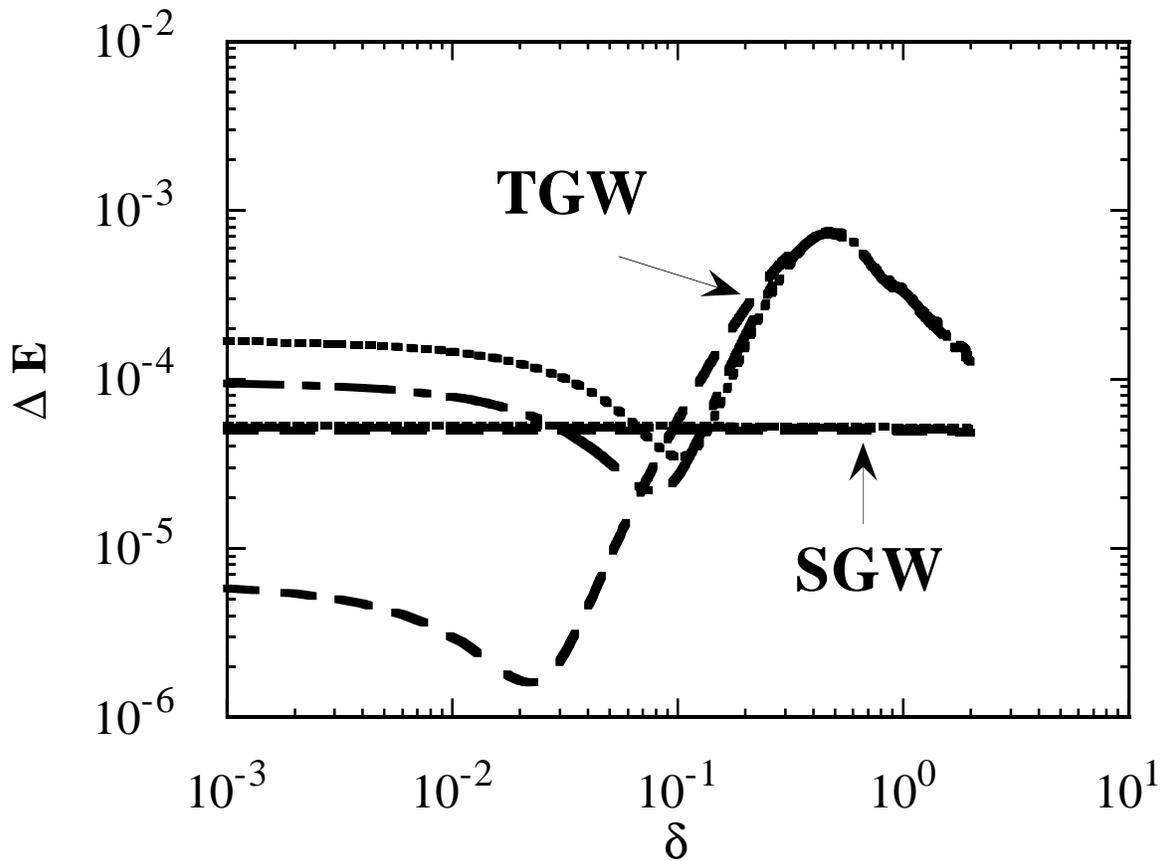
\end{document}